\documentclass[aps,prb,reprint,groupedaddress]{revtex4-1}

\usepackage{amsmath,amssymb} 
\usepackage{amsmath} 
\usepackage{amsfonts} 
\usepackage{bm} 
\usepackage[pdftex]{graphicx} 
\usepackage{braket} 

\begin{document}

\title{Discovering momentum-dependent magnon spin texture in insulating antiferromagnets: 
Role of the Kitaev interaction}

\author{Masataka Kawano}
\email{kawano@g.ecc.u-tokyo.ac.jp}
\author{Chisa Hotta}
\affiliation{Department of Basic Science, University of Tokyo, Meguro-ku, Tokyo 153-8902, Japan}
\date{\today}

\begin{abstract}
We theoretically show that the Kitaev interaction generates a novel class of spin texture in the excitation spectrum 
of the antiferromagnetic insulator found in the Kitaev-Heisenberg-$\Gamma$ model. 
In conducting electronic systems, there is a series of vortex type of spin texture along the Fermi surface
induced by Rashba and Dresselhaus spin-orbit couplings. 
Such spin textures are rarely found in magnetic insulators, 
since there had been no systematic ways to control the kinetics of its quasi-particle called magnon 
using a magnetic field or spacially asymmetric exchange couplings. 
Here, we propose a general framework to explore such spin textures in arbitrary insulating antiferromagnets.
We introduce an analytical method to transform any complicated Hamiltonian
to the simple representation based on pseudo-spin degrees of freedom,
which couples to the momentum-dependent fictitious ``Zeeman field''.
The direction of the pseudo-spin on a Bloch sphere describes
the degree of contributions from the two magnetic sublattices to the spin moment carried by the magnon.
There, the ``Zeeman field'' determines the direction of the pseudo-spin
and thus becomes the control parameter of the spin texture,
which is explicitly described by the original model parameters.
The framework enabled us to clarify the uncovered aspect of the Kitaev interaction, and further provides a tool to easily design or explore materials with intriguing magnetic properties. 
Since these spin textures can be a source of a pure spin current, 
the Kitaev materials $A_2$PrO$_3$ ($A=$Li, Na) shall become a potential platform 
of power-saving spintronics devices. 
\end{abstract}

\maketitle

\section{Introduction}
Exploring kaleidoscopic spin textures in crystalline solids is currently one of the most attractive topics in condensed matter physics, 
since it serves as the source of a wide variety of intriguing phenomena. 
In multiferroics, noncollinear spin textures such as spiral orders couple to an electric polarization,
which enables the control of the polarization by a magnetic field in TbMnO$_{3}$ \cite{kimura2003,katsura2005,sergienko2006,mostovoy2006}.
Noncoplanar spin textures, which are realized in skyrmion crystals MnSi \cite{muhlbauer2009,yu2010}
or pyrochlore magnets $A_{2}$Mo$_{2}$O$_{7}$ ($A$=Sm, Nd) \cite{taguchi1999,taguchi2001},
couple to electrons in a more sophisticated manner;
they produce an effective U(1) gauge field that works on an electron motion and generate a topological Hall effect
\cite{matl1998,ye1999,taguchi1999,ohgushi2000,taguchi2001,lyanda2001,neubauer2009,lee2009,kanazawa2011,li2013}. 
\par
Spin textures emerge also in momentum space as a consequence of electronic spin-orbit coupling (SOC). 
When a spatial-inversion symmetry is broken in bulk crystals or near the surface of materials, 
Rashba and Dresselhaus SOC's are typically induced, which give rise to momentum-dependent vortex and antivortex spin textures at the Fermi level
\cite{rashba1960,casella1960,bychkov1984,dresselhaus1955,dyakonov1986}.
These textures offer us an exciting playground to manipulate spins by an electric current,
e.g. by spin torques or spin-valves \cite{bernevig2005,manchon2008,chernyshov2009,miron2010,miron2011,kurebayashi2014},
a spin field-effect transistor \cite{datta1990,koo2009}. 
Also, when such spin-textured band is shifted by the current, the spin is polarized throughout the sample,
which is called inverse spin galvanic effect or Edelstein effect \cite{edelstein1990,kato2004,silov2004}.
However, the injection of electric currents is inextricably linked to Joule heating, which is unfavorable for device applications. 
Recently, there has been a growing interest in making use of magnetic insulators as replacements,
aiming to utilize the dissipationless spin current instead \cite{chumak2015}. 
\par
In magnetic insulators, the kinetic motion of electrons is lost. 
However, the effect of SOC still exists, as it is transformed into an antisymmetric exchange between localized spins, 
called Dzyaloshinskii-Moriya (DM) interaction \cite{dzyaloshinskii1958,moriya1960}. 
One essential feature of this interaction is that it affects the kinetic motion of magnons,
charge-neutral quasiparticles carrying spin-1. 
In insulating ferromagnets such as $A_{2}$V$_{2}$O$_{7}$ ($A$=Lu, Ho, In) \cite{onose2010,ideue2012} and Cu(1-3,bdc) \cite{hirschberger2015},
the DM interaction generates an effective U(1) gauge field,
which is transcribed into a Berry curvature in momentum space, and then becomes the origin of a thermal Hall effect 
\cite{fujimoto2009,katsura2010,onose2010,matsumoto2011prb,matsumoto2011prl,ideue2012,matsumoto2014,hirschberger2015}
and topologically protected surface magnons \cite{zhang2013,shindou2013,chisnell2015}.
\par
More recently, the role of DM interaction turned out to be far richer in insulating antiferromagnets;
there are two species of magnons belonging to two magnetic sublattices,
which can be regarded as ``up" and ``down" pseudo-spins in analogy with electronic spins.
Then, the DM interaction couples the pseudo-spins with the kinetic motion of magnons
in a similar manner to the SOC that couples the spins and the momentum of electrons.
Indeed, various tunable spin textures on magnon bands in momentum space are found
in 1D \cite{okuma2017} and square-lattice antiferromagnets \cite{kawano2019-1},
which are the models of Ba$_{2}X$Ge$_{2}$O$_{7}$ ($X$=Co, Mn) 
\cite{zheludev2003,kezsmarki2011,bordacs2012,penc2012,judit2012,masuda2010,murakawa2012,iguchi2018};
it can be regarded as a magnonic Rashba-Dresselhaus effect.
Also, on a honeycomb lattice, a spin Nernst effect occurs by the DM interaction
\cite{cheng2016,zyuzin2016,kovalev2016,shiomi2017,zhang2018,zyuzin2018} 
whose origin can be understood in a similar context. 
\par
The DM interaction is not the only outcome the SOC adds to magnetic insulators; 
it can be the source of a bond-dependent spin exchange called Kitaev interaction \cite{kitaev2006}. 
After the proposal that this interaction can be realized in Mott insulators with the strong SOC
\cite{jackeli2009,chaloupka2010}, the Kitaev models and related materials, 
iridium oxides $A_{2}$IrO$_{3}$ (A=Li, Na) \cite{singh2010,singh2012}
and ruthenium chloride $\alpha$-RuCl$_{3}$ \cite{plumb2014,kubota2015}, 
have been extensively studied. 
The Kitaev model is exactly solvable and is known to host a $\mathbb{Z}_{2}$ spin-liquid ground state, 
characterized by the fractionalization of spins into Majorana fermions and $\mathbb{Z}_{2}$ fluxes.
However, in the above-mentioned materials, there also exist a Heisenberg exchange interaction
and a so-called $\Gamma$-term, which together replace the expected spin-liquid phase with 
the magnetically-ordered phase at low temperatures in reality \cite{jackeli2009,chaloupka2010,rau2014,janssen2016}.
\par
In the present paper, we propose that this lowest-temperature state of the Kitaev materials can be more than just
a simple ordered antiferromagnet.
An exotic spin texture is found in its magnon excitation spectrum, whose origin is the spatially anisotropic Kitaev interaction.
This finding is made possible by the framework we propose together in this paper.
As a standard treatment for an arbitrary Hamiltonian of insulating antiferromagnets with a long-range magnetic order,
a bosonic Bogoliubov-de Gennes (BdG) Hamiltonian is derived, which describes a magnon excitation at low energies.
We construct a systematic way to \textit{exactly} transform this bosonic BdG Hamiltonian
to the representation based on the pseudo-spin degrees of freedom.
Classifying the symmetry of a pseudo-spin state immediately tells us \textit{analytically}
what parameter in the original model works to generate spin textures in what condition.
We demonstrate the usefulness of our framework, choosing the antiferromagnets in 1D and on a honeycomb lattice
with DM interactions as examples.
\par
Analyzing the excitation of the Kitaev model is not simple, in contrast to its ground state.
Nevertheless, we show analytically that the Kitaev interaction almost always generates intriguing spin textures,
and the $\Gamma$-term assists their variation in momentum space.
The framework also offers us information on which direction one needs to place the magnetic field 
to have a desiable spin texture, which can be utilized in experiments. 
Recently, the $f$-electron-based materials $A_{2}$PrO$_{3}$ ($A$=Li, Na) \cite{jang2019}
and ruthenium trihalides with multiple anions $\alpha$-RuH$_{3/2}$X$_{3/2}$ (X=Cl, Br) \cite{sugita2019}
are theoretically proposed. 
These materials host antiferromagnetic Kitaev exchange interaction,
and on the top of that, $A_{2}$PrO$_{3}$ has the antiferromagnetic Heisenberg exchange interaction,
which stabilizes a N\'{e}el order in the ground state.
Our results are directly applied to these materials, 
and further offers a chance of finding a more abundant platform of the physics of spin textures
in insulating antiferromagnets, which is now in quest. 
\par
The paper is organized as follows.
In Sec. \ref{sec:formulation}, we first present our theoretical framework 
by using the analogy with electronic systems. 
The details and proofs of the formulation are given in Appendices \ref{sec:bdg}-\ref{sec:app1}.
In Sec. \ref{sec:model}, we apply our framework to three types of antiferromagnets,
a 1D antiferromagnet, Kitaev-Heisenberg-$\Gamma$ model, and a honeycomb-lattice antiferromagnet,
showing that a variety of spin textures can be actually explored and classified.
We finally give a brief summary and discussions in Sec. \ref{sec:conclusion}.

\section{Spin texture in momentum space}
\label{sec:formulation}
\subsection{Electrons in metals}
\label{subsec:overview}
\par
As a prototype reference, we first show how the spin textures are formed in momentum space 
in the case of electronic systems with SOC. 
The Bloch Hamiltonian of electrons in solids with a single orbital per sites can be written as
\begin{equation}
H(\bm{k})=
R^{0}(\bm{k})\sigma^{0}+\bm{R}(\bm{k})\cdot\bm{\sigma},
\label{eq:H_electron}
\end{equation}
where $\bm{R}(\bm{k})=(R^{x}(\bm{k}),R^{y}(\bm{k}),R^{z}(\bm{k}))$ and $R^{\mu}(\bm{k})\in\mathbb{R}$ ($\mu=0,x,y,z$),
$\sigma^{0}$ is the unit matrix,
and $\bm{\sigma}=(\sigma^{x},\sigma^{y},\sigma^{z})$ is the Pauli matrix representing the electron spin degrees of freedom.
In the absence of SOC, the up and down electron spins do not couple so that $\bm{R}(\bm{k})=\bm{0}$.
\par
\begin{figure*}[tbp]
\includegraphics[width=180mm]{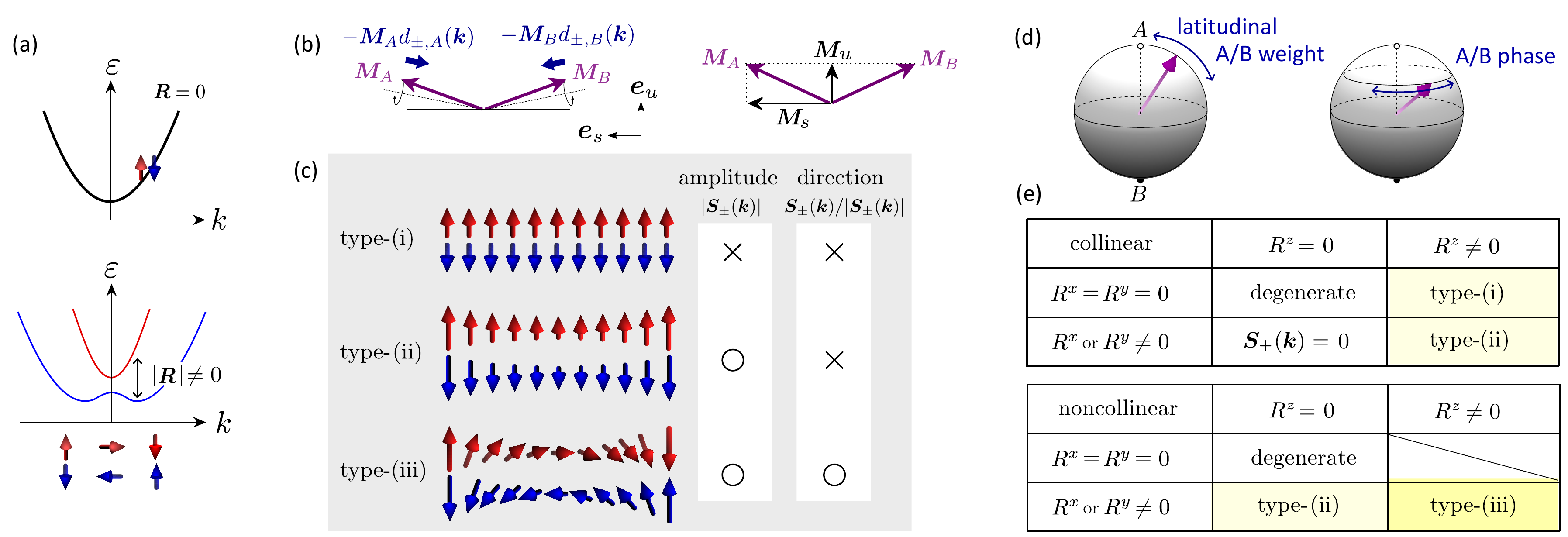}
\caption{(a) Schematic illustration of energy bands,
$\varepsilon_{\pm}^{(\mathrm{el})}(\bm{k})=R_{0}(\bm{k})\pm|\bm{R}(\bm{k})|$
for $\bm R=0$ and $\bm R\ne 0$. $\bm R(\bm k)$ works as a $\bm k$-dependent ``Zeeman field". 
(b) Antiferromagnetic moments, $\bm{M}_{A/B}$, of the ground state on two sublattices. 
On each lattice site, the sublattice magnetization shrinks by $-\bm{M}_{A/B} d_{\pm,A/B}(\bm{k})$, 
where $d_{\pm,A/B}(\bm{k})$ represents the local spectral weight of magnon bands. 
The unit vectors, $\bm{e}_{u}$ and $\bm{e}_{s}$, describe the directions of the uniform and staggered 
directions of the magnetic moments, which $\bm M_u$ and $\bm M_s$ point to. 
(c) Classification of magnon spin textures (i)--(iii) based 
on the $\bm k$-dependence of the amplitude and the direction of $\bm{S}_{\pm}(\bm{k})$, 
where the upper/lower arrows denote $\bm{S}_{+}(\bm{k})/\bm{S}_{-}(\bm{k})$ belonging to the upper/lower bands. 
(d) Direction of the pseudo-spin on a Bloch sphere, which is parallel to $\bm{R}(\bm{k},\varepsilon)$ in Eq. (\ref{eq:heff}). 
North and south poles represent the $d_{n,A}(\bm{k})-d_{n,B}(\bm{k})=+1$ and $-1$ states, respectively,
and the equator gives $d_{n,A}(\bm{k})=d_{n,B}(\bm{k})$ ($n=\pm$).
In varying $R^{z}(\bm{k},\varepsilon)/|\bm{R}(\bm{k},\varepsilon)|$, the pseudo-spin moves along the latitudinal direction,
and when $R^{x}(\bm k,\varepsilon)/R^{y}(\bm k,\varepsilon)$  changes,
the pseudo-spin rotates about the $z$-axis.
(e) The relationships between $\bm{R}(\bm{k},\varepsilon)$ and the magnon spin textures.
The noncollinear spin configuration with $R^{z}(\bm{k},\varepsilon)\neq0$ leads to type-(iii) magnon spin textures.}
\label{fig1}
\end{figure*}

By solving the eigenvalue equation of $H(\bm{k})$,
\begin{equation}
H(\bm{k})\bm{w}_{\pm}^{(\mathrm{el})}(\bm{k})\
=\varepsilon_{\pm}^{(\mathrm{el})}(\bm{k})\bm{w}_{\pm}^{(\mathrm{el})}(\bm{k})\
,
\label{eq:H_secular}
\end{equation}
we obtain the energy bands $\varepsilon_{\pm}^{(\mathrm{el})}(\bm{k})=R^{0}(\bm{k})\pm|\bm{R}(\bm{k})|$. 
The two bands split when $|\bm{R}(\bm{k})|\neq 0$ as shown in Fig. \ref{fig1}(a). 
The degree of splitting depends on $\bm{k}$, and $\bm{R}(\bm{k})$ can be
regarded as an effective $\bm{k}$-dependent ``Zeeman field".
This field points in the $\hat{\bm{R}}(\bm{k})=\bm{R}(\bm{k})/|\bm{R}(\bm{k})|$ direction, so that
the upper and lower bands carry the spins that are pointing in the $\pm\hat{\bm{R}}(\bm{k})$ directions, 
which is indeed evaluated for each eigenstate $\bm{w}_{\pm}^{(\mathrm{el})}(\bm{k})$ as 
\begin{equation}
\bm{S}_{\pm}^{(\mathrm{el})}(\bm{k})
=
\frac{1}{2}\{\bm{w}_{\pm}^{(\mathrm{el})}(\bm{k})\}^{\dagger}\bm{\sigma}\bm{w}_{\pm}^{(\mathrm{el})}(\bm{k})
=
\pm\frac{1}{2}\hat{\bm{R}}(\bm{k})
.
\label{eq:S_el}
\end{equation}
A momentum-dependent spin texture thus emerges when the direction of the ``Zeeman field'' varies with $\bm{k}$:
\begin{equation}
\hat{\bm{R}}(\bm{k})\neq\mathrm{const}
. 
\label{eq:R_condition_el}
\end{equation}
Indeed the Hamiltonian including the Rashba and Dresselhaus SOC terms gives 
$\bm R(\bm k)=(-\alpha k_y, \alpha k_x,0)$ and $(-\beta k_x, \beta k_y,0)$, respectively, 
with the coupling constant $\alpha$ and $\beta$.
Since $\bm{R}(\bm{k})$ rotates with $\bm{k}$ in the $xy$-plane,
the vortex/anti-vortex type of spin texture emerges in momentum space
\cite{rashba1960,casella1960,bychkov1984,dresselhaus1955,dyakonov1986}. 
%
\par
In the multi-orbital systems, the Bloch Hamiltonian is described by a $2n\times2n$ ($n>2$) Hermitian matrix.
Therefore, the simple picture about the coupling of spins with momentum, obtained in Eq. (\ref{eq:H_electron}) no longer applies.
However, such seemingly complicated Hamiltonian can be formally reduced to the simple form of Eq. (\ref{eq:H_electron}),
if we adopt the Brillouin-Wigner formalism.
This formalism was used to clarify the analytical condition for Dirac band dispersions in electronic systems \cite{asano2011}.
For later convenience, we consider the case of $n=2$,
whose Hamiltonian and its retarded Green function can be generally written as
\begin{equation}
H(\bm{k})
=
\sum_{\mu,\nu=0,x,y,z}h_{\mu\nu}(\bm{k})\tau^{\mu}\otimes\sigma^{\nu}
=
\begin{pmatrix}
H_{11}(\bm{k}) & H_{12}(\bm{k})\\
H_{21}(\bm{k}) & H_{22}(\bm{k})
\end{pmatrix}
,
\label{eq:H_orbital}
\end{equation}
\begin{equation}
G^{(\mathrm{el})}(\bm{k},\varepsilon)
=
[\varepsilon\tau^{0}\otimes\sigma^{0}-H(\bm{k})]^{-1}
=
\begin{pmatrix}
G_{11}(\bm{k},\varepsilon) & G_{12}(\bm{k},\varepsilon)\\
G_{21}(\bm{k},\varepsilon) & G_{22}(\bm{k},\varepsilon)
\end{pmatrix}
,
\label{eq:G_orbital}
\end{equation}
where $h_{\mu\nu}(\bm{k})$, $\varepsilon\in\mathbb{R}$ 
and $\tau^{\mu}$ ($\mu=0,x,y,z$) denotes the unit and Pauli matrices acting on the orbital space.
$H_{ll'}(\bm{k})$ and $G_{ll'}(\bm{k},\varepsilon)$ are the $2\times2$ matrices, and $l=1,2$ is the orbital index.
The pole and residue of $G^{(\mathrm{el})}(\bm{k},\varepsilon)$ give all the eigenvalues and eigenstates of $H(\bm{k})$. 
The Green function in the subspace of the $l=1$ orbital, $G_{11}(\bm{k},\varepsilon)$,
also gives all the eigenvalue and eigenstates projected onto the $l=1$ subspace.
Thus, we can obtain the effective Hamiltonian for $l=1$ orbital as
\begin{equation}
G_{11}(\bm{k},\varepsilon)
=
[\varepsilon\sigma^{0}-H_{\mathrm{eff}}^{l=1}(\bm{k},\varepsilon)]^{-1}
,
\label{eq:G_11}
\end{equation}
\begin{align}
H_{\mathrm{eff}}^{l=1}(\bm{k},\varepsilon)
&=
H_{11}(\bm{k})
+
H_{12}(\bm{k})[\varepsilon\sigma^{0}-H_{22}(\bm{k})]^{-1}H_{21}(\bm{k})
\nonumber \\
&=
R_{l=1}^{0}(\bm{k},\varepsilon)\sigma^{0}
+
\bm{R}_{l=1}(\bm{k},\varepsilon)\cdot\bm{\sigma}
.
\label{eq:H_eff_electron}
\end{align}
One can formally solve the eigenvalue equation of Eq. (\ref{eq:H_eff_electron}) as we did in Eq. (\ref{eq:H_secular}),
and the solution is obtained in the following form,
\begin{equation}
\varepsilon=R_{l=1}^{0}(\bm{k},\varepsilon)\pm|\bm{R}_{l=1}(\bm{k},\varepsilon)|,  
\label{eq:obtain_en}
\end{equation}
where both sides of the equation include the parameter $\varepsilon$,
that takes the same value in principle.
The vector $\bm{R}_{l=1}(\bm{k},\varepsilon_{n}^{(\mathrm{el})}(\bm{k}))$ represents the contribution
from the $l=1$ orbital to the electron spin at the band-$n$. 
The actual value of the energy $\varepsilon=\varepsilon_{n}^{(\mathrm{el})}(\bm{k})$
is obtained by solving Eq.(\ref{eq:obtain_en}), 
which is equivalent to obtaining the eigenvalue of Eq.(\ref{eq:H_orbital}).
However, even by keeping $\varepsilon$ unknown,
one can formally regard $\bm{R}_{l}(\bm{k},\varepsilon)$ as an effective $\bm k$-dependent ``Zeeman field''
that splits the energy bands carrying spins come from the $l$-orbital, pointing in the directions, 
$\pm \hat{\bm{R}}_{l}(\bm{k},\varepsilon_{n}^{(\mathrm{el})}(\bm{k}))$.
The necessary condition to have a momentum-dependent spin texture is given by
$\hat{\bm{R}}_{l}(\bm{k},\varepsilon_{n}(\bm{k}))\neq\mathrm{const}$.

\subsection{Insulating antiferromagnets}
\subsubsection{Bosonic BdG Hamiltonian}
\label{subsec:BdG}
Let us consider a generic two-sublattice antiferromagnetic insulator, 
whose Hamiltonian $\hat{\mathcal{H}}$ is written in terms of a spin operator $\hat{\bm{S}}_{i}$ on a site $i$, 
e.g. a Heisenberg model, XXZ model, and so on. 
We confine ourselves to the case where the system has a long-range magnetic order in the ground state. 
The magnetic unit cell consists of two sublattices, $A$ and $B$, and the directions of the classical spins
on these sublattices are given by the unit vectors denoted as $\bm{M}_{A}$ and $\bm{M}_{B}$, 
which are determined in advance so as to minimize the energy of the classical ground state of the Hamiltonian.
In the spin-wave theory, the quantum fluctuation that represents the excitation from this ground state
is described by the Holstein-Primakoff bosons $\hat{a}_{i}/\hat{b}_{i}$ defined on the sites belonging 
to the sublattice $A/B$. 
The spin moment at a site $i\in A$ shrinks as 
\begin{align}
\braket{\hat{S}_{i}}
&=
\left(
S-\braket{\hat{a}_{i}^{\dagger}\hat{a}_{i}}
\right)
\bm{M}_{A}
,
\label{eq:S-shrink-new}
\end{align}
where $S$ is the spin quantum number and $\langle \cdots \rangle$ is the expectation value 
in terms of some target excited state or a mixed state at finite temperature. 
The expectation value of the spin at a site $i\in B$ can also be obtained in a similar manner.
\par
The effective low-energy Hamiltonian takes the quadratic form,
$\hat{\mathcal{H}}\simeq\hat{\mathcal{H}}_{\mathrm{sw}}+\mathrm{const}$,
with $\hat{\mathcal{H}}_{\mathrm{sw}}$ given by
\begin{equation}
\hat{\mathcal{H}}_{\mathrm{sw}}=
\frac{1}{2}\sum_{\bm{k}}
\hat{\Phi}^{\dagger}(\bm{k})
H_{\mathrm{BdG}}(\bm{k})
\hat{\Phi}(\bm{k})
,
\label{eq:ham_sw}
\end{equation}
where $\hat{\Phi}(\bm{k})=(\hat{a}_{\bm{k}},\hat{b}_{\bm{k}},\hat{a}^{\dagger}_{-\bm{k}},\hat{b}^{\dagger}_{-\bm{k}})^{T}$, 
and $\hat{a}_{\bm{k}}/\hat{b}_{\bm{k}}$ is the Fourier transform of $\hat{a}_{i}$/$\hat{b}_{i}$. 
The bosonic BdG Hamiltonian $H_{\mathrm{BdG}}(\bm{k})$ is the $4\times 4$ Hermitian matrix, 
and it can be generally expressed as 
\begin{equation}
H_{\mathrm{BdG}}(\bm{k})=
\begin{pmatrix}
\Xi(\bm{k}) & \Delta(\bm{k})\\
\Delta^{*}(-\bm{k}) & \Xi^{*}(-\bm{k})
\end{pmatrix}
,
\label{eq:H_BdG}
\end{equation}
where $\Xi(\bm{k})$ and $\Delta(\bm{k})$ are the $2\times 2$ matrices 
satisfying $\Xi^{\dagger}(\bm{k})=\Xi(\bm{k})$ and $\Delta^{\dagger}(\bm{k})=\Delta^{*}(-\bm{k})$. 
Magnon bands $\varepsilon_{\pm}(\bm{k})$ and corresponding eigenvectors 
$\bm{t}_{\pm}(\bm{k})=(u_{\pm, A}(\bm{k}), u_{\pm,B}(\bm{k}),v_{\pm, A}(\bm{k}), v_{\pm,B}(\bm{k}))^{T}$
can be obtained by solving the eigenvalue equation of the \textit{non-Hermitian} matrix $\Sigma^{z}H_{\mathrm{BdG}}(\bm{k})$
with $\Sigma^{z}=\tau^{z}\otimes\sigma^{0}$, where $\tau^{\mu}$ and $\sigma^{\mu}$ are the unit and Pauli matrices
acting on a particle-hole and sublattice space, respectively (see Appendix \ref{sec:bdg}). 
The local spectral weight, namely the local density of the eigenstates on A and B sublattices, is given as
\begin{equation}
d_{\pm,A/B}(\bm{k})
=
|u_{\pm,A/B}(\bm{k})|^{2}
+
|v_{\pm,A/B}(\bm{k})|^{2}
\label{eq:local_spectral_weight}
.
\end{equation}
\par
At finite temperature, the moments in Eq.(\ref{eq:S-shrink-new}) is rewritten as
\begin{equation}
\braket{\hat{S}_{i}}
=\!
\Big(\!
S-\delta S_{A}
-\frac{1}{N_{c}}\sum_{\bm{k},n=\pm}g(\varepsilon_{n}(\bm{k}))d_{n,A}(\bm{k})
\!\Big)\!
\bm{M}_{A}
,
\label{eq:S-shrink}
\end{equation}
where $\delta S_{A}=(1/N_{c})\sum_{\bm{k},n}|v_{n,A}(\bm{k})|^{2}$ denotes the contribution from a zero-point fluctuation,
$N_{c}$ is the number of unit cells in the system
and $g(\varepsilon)$ is the Bose distribution function.
One can see that the local spectral weight $d_{n,A/B}(\bm{k})$ denotes the amplitude of how much the 
eigenstate of magnon bands $n=\pm$ at $\bm{k}$ 
contribute to the shrinking of ordered moments on the sublattice $A/B$. 
The spin moments in momentum space carried by magnons
are thus the ones that suppress the spin moments in real space from those of the ground state
as $-\bm{M}_{A/B}d_{\pm,A/B}(\bm{k})$, as shown 
schematically in Fig.~\ref{fig1}(b). 
The ones on upper ($+$) and lower ($-$) bands are defined as \cite{okuma2017}
\begin{align}
\bm{S}_{\pm}(\bm{k})
&=
-\bm{M}_{A}d_{\pm,A}(\bm{k})
-\bm{M}_{B}d_{\pm,B}(\bm{k})
\\
&=
-\bm{M}_{u}\left(d_{\pm,A}(\bm{k})+d_{\pm,B}(\bm{k})\right)
\nonumber \\
&\hspace{60pt}
-\bm{M}_{s}\left(d_{\pm,A}(\bm{k})-d_{\pm,B}(\bm{k})\right)
,
\label{eq:S_magnon}
\end{align}
where we introduce the magnetic moments in the uniform direction, $\bm{M}_{u} \equiv (\bm M_{A} + \bm M_{B})/2$, 
and staggered direction, $\bm{M}_{s} \equiv (\bm M_{A} - \bm M_{B})/2$, for later convenience. 
The vector $\bm{S}_{\pm}(\bm{k})$ is defined on each $\bm{k}$-point in the Brillouin zone,
and forms magnon spin textures in momentum space. 
The straightforward treatment is to solve the eigenvalue equation of $H_{\rm BdG}(\bm{k})$ directly and obtain $d_{\pm,A/B}(\bm{k})$.
However, it is often difficult to analytically obtain $d_{\pm,A/B}(\bm{k})$ except for some simple cases.
Even if the spin textures are feasible, one often needs to fine-tune numbers of parameters,
such as the direction of the magnetic field and the way of alignment of DM vectors.
We thus provide a systematic and efficient framework to judge how and when the spin textures emerge in insulating antiferromagnets.

\subsubsection{Pseudo-spin degrees of freedom}
\label{subsec:BW}
The central idea of our framework is to extract an effective ``Zeeman field'', 
$\bm R(\bm k,\varepsilon)$, that couples to the pseudo-spin degrees of freedom 
from the bosonic BdG Hamiltonian using the Brillouin-Wigner formalism. 
We only outline the results in the following for the sake of clarity.
Although they look simple enough, the derivation and proofs for the bosonic case are not as straightforward; 
since the magnons are bosons, their Green's functions are the \textit{non-Hermitian} matrices, 
and their eigenstates are those of the \textit{non-Hermitian} matrix $\Sigma^{z}H_{\mathrm{BdG}}(\bm{k})$.
The proofs and the details of applying the Brillouin-Wigner formalism, 
reducing the $4\times4$ \textit{non-Hermitian} matrix to a $2\times2$ \textit{Hermitian} one
are given in Appendices. \ref{subsec:H_eff} and \ref{sec:app1}. 
\par
We focus on the Green function in a particle subspace, 
\begin{equation}
G(\bm{k},\varepsilon)
=
[\varepsilon\sigma^{0}-H_{\mathrm{eff}}(\bm{k},\varepsilon)]^{-1}
,
\label{eq:gfunc}
\end{equation}
where the contribution from the pair creation-annihilation terms $\Delta(\bm{k})$, are renormalized 
into the effective Hamiltonian
\begin{align}
H_{\mathrm{eff}}(\bm{k},\varepsilon)
&=\Xi(\bm{k})-\Delta(\bm{k})[\varepsilon\sigma^{0}+\Xi^{*}(-\bm{k})]^{-1}\Delta^{\dagger}(\bm{k})
\nonumber \\
&=R^{0}(\bm{k},\varepsilon)\sigma^{0}+\bm{R}(\bm{k},\varepsilon)\cdot\bm{\sigma},
\label{eq:heff}
\end{align}
which is the $2\times2$ Hermitian matrix, namely it has the same form as the fermionic (electronic) ones. 
The unit and Pauli matrices $\sigma^{\mu}$ ($\mu=0,x,y,z$) act on the sublattice space.
The effective Hamiltonian, namely the pole and residue of Eq.(\ref{eq:gfunc}), 
\textit{exactly} reproduces the magnon bands and eigenstates of the bosonic BdG Hamiltonian,
and we find an analytical relationship between $d_{\pm,A/B}(\bm{k})$ and $\bm{R}(\bm{k},\varepsilon)$
(see Appendix. \ref{subsec:H_eff}).
\par
The key picture presented in Eq.(\ref{eq:heff}) is that the $A$ and $B$ sublattices degrees of freedom
are regarded as pseudo-spin degrees of freedom.
For example, if this pseudo-spin has an SU(2) symmetry,
\begin{equation}
[H_{\rm eff}(\bm{k},\varepsilon),\sigma^{\mu}]=0,
\hspace{20pt}
(\mu=x,y,z)
,
\label{eq:U1_H_eff}
\end{equation}
the ``Zeeman field'' vanishes, $\bm{R}(\bm{k},\varepsilon)=\bm{0}$.
This context is equivalent to the case of electrons in Eq.(\ref{eq:H_electron});
the SU(2) symmetry of spins is kept when $\bm{R}(\bm{k})=\bm{0}$, in which case the up and down spin bands are fully degenerate
(see Appendix \ref{subsec:symmetry} for details).
\par
Despite the above-mentioned similarity, 
the way how the spin textures are generated in the case of this antiferro-magnons differs from that of electrons.
To clarify this point, we examine the role of the ``Zeeman field'' and the pseudo-spin.
Since the pseudo-spin points in the direction parallel/antiparallel to $\bm{R}(\bm{k},\varepsilon)$, 
its direction may vary with $\bm{k}$. 
Notice that the direction of pseudo-spin indicates the relative weight and phase of
$\hat{a}$- and $\hat{b}$-magnons that contribute to the eigenstate at the $\bm{k}$-point.
One can introduce a geometrical representation known as a Bloch sphere as shown in Fig.\ref{fig1}(d).
If the arrow representing the pseudo-spin points in the $+z$-direction, the state consists only of $\hat{a}$-magnons,
and if it points to the equator, the two species of magnons mix with equal weights. 
Therefore, the $z$-component of the ``Zeeman field'' $R^{z}(\bm{k},\varepsilon)$ 
determines the relative weight of magnons on the sublattice $A/B$. 
\par
The pseudo-spin is not the spin moment itself. 
Once the direction of the pseudo-spin is given, 
the weights of the $\hat{a}/\hat{b}$-magnons are fixed, each carrying the spin moments in the $-\bm M_{A/B}$ direction 
(see Eq.(\ref{eq:S_magnon})). 
Therefore, spin textures depend on whether $\bm M_{A}$ and $\bm M_{B}$ are collinear or noncollinear.
\par
Before discussing the relationships between $\bm{R}(\bm{k},\varepsilon)$ and $\bm{S}_{\pm}(\bm{k})$, 
we summarize in Fig. \ref{fig1}(c) the classification of three different spin textures \cite{kawano2019-1}:
(i) neither the direction nor the amplitude of spins vary with $\bm{k}$, 
(ii) the amplitude of spins varies with $\bm{k}$ but the direction does not,
and
(iii) both the direction and the amplitude vary with $\bm{k}$. 
Only type-(iii) shows the rich directional variation of $\bm{S}_{\pm}(\bm{k})$
\footnote{More precisely, there is also the type where the direction of spins varies with $\bm{k}$, but the amplitude does not. This type of spin textures can be seen in electronic systems, but generally cannot be found in magnonic ones. Then we does not consider this type.},
which we would like to search for. 
\par
Let us start from the case of collinear antiferromagnets, $\bm M_{A}=-\bm M_{B} = \bm M_s$, and $\bm M_u=0$, 
which leads to $\bm{S}_{\pm}(\bm{k})=-\bm{M}_s(d_{\pm,A}(\bm{k})-d_{\pm,B}(\bm{k}))$ in Eq.(\ref{eq:S_magnon}). 
When $R^{z}(\bm{k},\varepsilon)=0$, the pseudo-spin points to the equator, which gives 
$d_{\pm,A}(\bm{k})=d_{\pm,B}(\bm{k})$ and $\bm{S}_{\pm}(\bm{k})=0$.
When $R^{z}(\bm{k},\varepsilon)\neq0$ and $R^{x}(\bm{k},\varepsilon)=R^{y}(\bm{k},\varepsilon)=0$, 
the pseudo-spin points toward the north or the south pole, 
and we find $|d_{\pm,A}(\bm{k})-d_{\pm,B}(\bm{k})|=1$. 
Then the spin moment is locked to $\pm\bm{M}_{s}$ for the two bands, which is type-(i).
Otherwise, when $R^{z}(\bm{k},\varepsilon)\neq0$ while $R^{x}(\bm{k},\varepsilon)$ or $R^{y}(\bm{k},\varepsilon)$ varies with $\bm{k}$,
the pseudo-spin moves along the latitudinal direction, leading to $d_{A,\pm}(\bm{k})-d_{B,\pm}(\bm{k})\neq\mathrm{const}$, 
which is type-(ii).
\par 
For the noncollinear case, both $\bm{M}_{u}$ and $\bm{M}_{s}$ start to contribute to $\bm{S}_{\pm}(\bm{k})$. 
$\bm M_u$ couples to $d_{\pm,A}(\bm{k})+d_{\pm,B}(\bm{k})$ 
which generally varies with $\bm{k}$ and stretches the amplitude $|\bm{S}_{\pm}(\bm{k})|$. 
$\bm{M}_{s}$ couples to $d_{\pm,A}(\bm{k})-d_{\pm,B}(\bm{k})$ which is zero 
when $R^{z}(\bm{k},\varepsilon)=0$ since the pseudo-spins point to the equator, 
and we find the fixed direction, $\bm{S}_{\pm}(\bm{k})\parallel\bm{M}_{u}$, 
leading to type-(ii). 
When $R^{z}(\bm{k},\varepsilon)\neq0$, $\bm{S}_{\pm}(\bm{k})$ varies in the $\bm{M}_{s}$-direction 
by the latitudinal motion of the pseudo-spins, and together with the contribution from the $\bm{M}_{u}$-direction, 
form a type-(iii) spin texture
\footnote{Generally, a noncollinear spin configuration leads to a finite $R^{x}(\bm{k},\varepsilon)$ or $R^{y}(\bm{k},\varepsilon)$
except for the models where the $A$ and $B$ sublattices are completely decoupled.
Also $R^{x}(\bm{k},\varepsilon)$, $R^{y}(\bm{k},\varepsilon)$ and $R^{z}(\bm{k},\varepsilon)$ have different $\bm{k}$-dependence except for fine-tuned models. Then $R^{z}(\bm{k},\varepsilon)/|\bm{R}(\bm{k},\varepsilon)|$ varies with $\bm{k}$ when $R^{x}(\bm{k},\varepsilon)\neq0$ and $R^{z}(\bm{k},\varepsilon)\neq0$ for example.}.
\par
We would like to stress here that
one generally finds $d_{\pm,A}(\bm{k})+d_{\pm,B}(\bm{k})\neq1$, which is the particular feature of magnons as bosonic particles.
Contrastingly, in the electronic systems, the spectral weight of each one-body discrete energy level is always equal to 1.
Namely there is no ``stretching mode'' of spin moments, and only the directional variation of spins are allowed to exist
as one can see from the cases of Rashba and Dresselhaus electrons.
Therefore, type-(ii) and (iii) spin textures are the characteristic phenomena found in the antiferromagnetic insulator.
\par
To summarize this section, 
once the bosonic BdG Hamiltonian Eq.(\ref{eq:H_BdG}) is obtained by the standard treatment from the original spin Hamiltonian,
by deriving the analytical form of $\bm{R}(\bm{k},\varepsilon)$ from Eq. (\ref{eq:heff}),
and using the Tables in Fig. \ref{fig1}(e),
one can easily judge what parameters in the original spin Hamiltonian works to generate what types of spin textures.
Using this framework, it is also possible to design material systems that generate a desirable spin texture.
So far, type-(i) and (ii) are observed in many ferro or antiferromagnets,
but type-(iii) is observed only in the 1D and 2D antiferromagnets with DM interactions
\footnote{While there is a demonstration on the kagome lattice ferromagnets in Ref. \cite{okuma2017}, its ground state is not an ordered magnet both in the classical and the quantum cases, so that the spin-wave theory does not apply.}.
In the following section, we show some unprecedented example; in a KH$\Gamma$ model,
the Kitaev interaction and the $\Gamma$-term cooperatively generate the type-(iii) magnon spin texture.
%
%
\section{Application to models}
\label{sec:model}
In this section, we first apply our framework to the 1D antiferromagnet with DM interaction
to demonstrate that the pseudo-spin picture based on the two sublattice degrees of freedom 
is useful to classify magnon spin textures in momentum space. 
We then examine the details of a peculiar magnon spin texture in the KH$\Gamma$ model
and a honeycomb-lattice antiferromagnet with DM interaction, 
showing that our framework can easily determine what types of spin textures emerge
and what parameters play a crucial role even for complicated and nontrivial models. 
%
%
\subsection{1D antiferromagnet}
\begin{figure}[tbp]
\includegraphics[width=85mm]{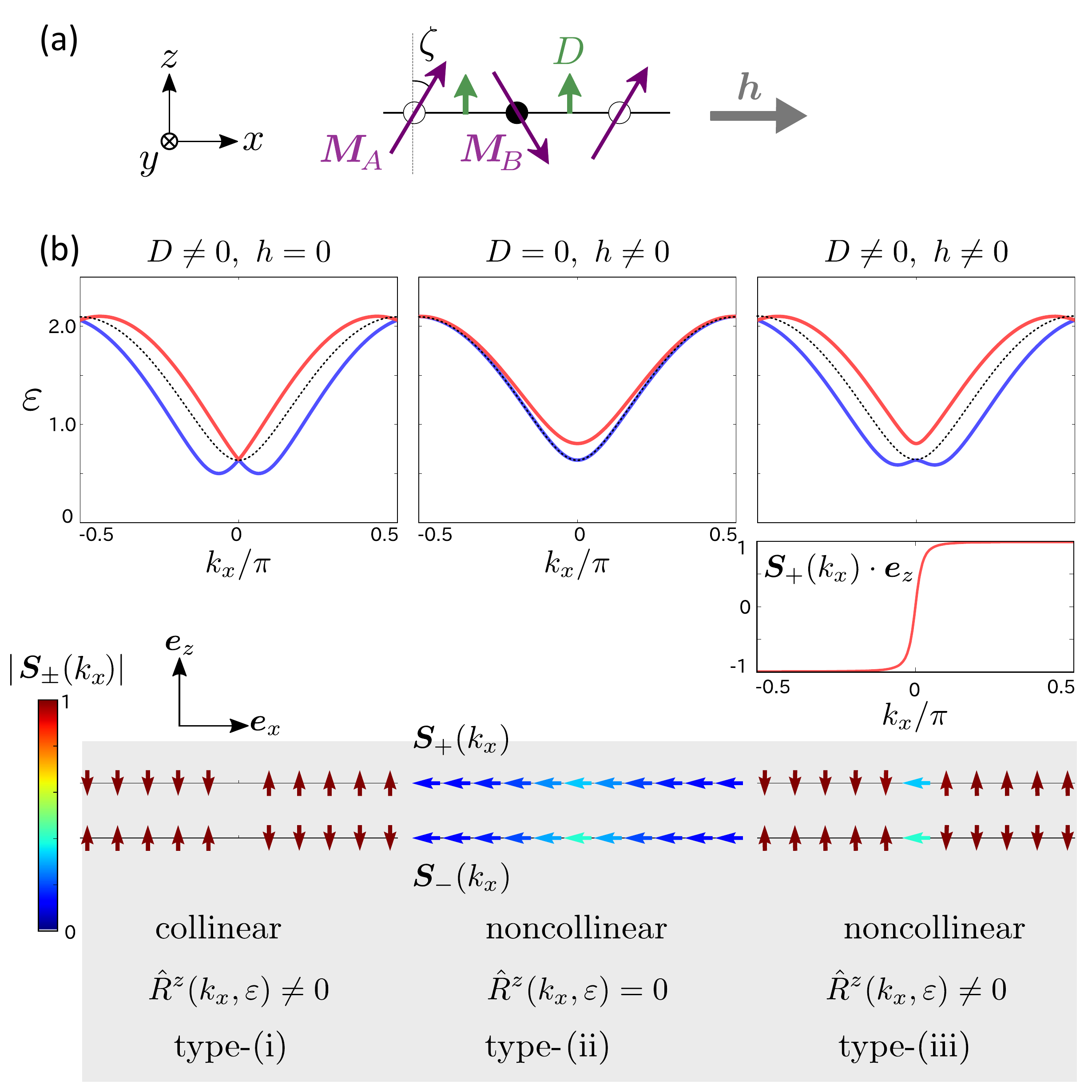}
\caption{(a) 1D antiferromagnet.
Purple, green and gray arrows indicate the spin configuration in the classical ground state,
the DM vector, and the magnetic field, respectively.
(b) Magnon bands and spin textures for
$D=0.2$ and $h=0$, $D=0$ and $h=0.5$, and $D=0.2$ and $h=0.5$.
The dotted line denotes the doubly degenerate magnon bands for $D=h=0$.
We set other parameters as $S=1.0$, $J=1.0$, and $\Lambda=0.05$.
The direction of spins is depicted by taking $\bm{e}_{x}$ and $\bm{e}_{z}$ as the basis,
where $\bm{e}_{\mu}$ is the unit vector pointing the $\mu$-spin axis ($\mu=x,y,z)$.
The noncollinear configuration and the finite $R^{x}(\bm{k},\varepsilon)$ and $R^{z}(\bm{k},\varepsilon)$
lead to the type-(iii) magnon spin texture.
}
\label{fig2}
\end{figure}

We consider the 1D antiferromagnet in Fig.\ref{fig2}(a), described by the following Hamiltonian:
\begin{align}
\hat{\mathcal{H}}
&=
J
\sum_{j}
\hat{\bm{S}}_{j}
\cdot
\hat{\bm{S}}_{j+1}
+
D
\sum_{j}
\bm{e}_{z}
\cdot
(
\hat{\bm{S}}_{j}
\times
\hat{\bm{S}}_{j+1}
)
\nonumber \\
&\hspace{20pt}
-
\Lambda
\sum_{j}
(\hat{S}_{j}^{z})^{2}
-
h
\sum_{j}
\hat{S}_{j}^{x}
,
\end{align}
where $J$ ($>0$) is the Heisenberg exchange interaction,
$\Lambda$ ($>0$) is the easy-axis anisotropy,
$D$ is the Dzyaloshinskii-Moriya interaction,
and $h$ denotes the magnetic field pointing in the $x$-direction.
This type of antiferromagnet has been studied
in the context of a nonreciprocity \cite{hayami2016,gitgeatpong2017},
device applications \cite{cheng2016-1D},
and a magnon spin-momentum locking \cite{okuma2017}.
For sufficiently small $D$,
a canted N\'eel order is realized in a classical ground state (see Fig. \ref{fig2}(a)),
where $\Lambda$ plays a role to suppress the spiral order to keep the canted antiferromagnetic structure.
The canting angle is given by
$\zeta=\arcsin\big(h/(2(2J+\Lambda)S)\big)$.
\par
By applying the linear spin-wave theory, 
we obtain $\Xi(\bm{k})$ and $\Delta(\bm{k})$ in Eq. (\ref{eq:H_BdG}) as
\begin{align}
\Xi(k_{x})
&=
\Xi_{0}\sigma^{0}
+
\Xi_{x}(k_{x})\sigma^{x}
,
\\
\Delta(k_{x})
&=
\Delta_{0}\sigma^{0}
+
\Delta_{x}(k_{x})\sigma^{x}
+
\Delta_{y}(k_{x})\sigma^{y}
,
\end{align}
with
\begin{align}
&\Xi_{0}=
2JS+\Lambda S(1+\cos^{2}\zeta)
,
\nonumber\\
&\Xi_{x}(k_{x})=
2JS\sin^{2}\zeta\cos k_{x}
,
\nonumber\\
&\Delta_{0}=
-\Lambda S\sin^{2}\zeta
,
\nonumber\\
&\Delta_{x}(k_{x})=
-2JS\cos^{2}\zeta\cos k_{x}
,
\nonumber\\
&\Delta_{y}(k_{x})=
-2iDS\cos\zeta\sin k_{x}
.
\label{eq:dy}
\end{align}
Then, the ``Zeeman field" is calculated from Eq.(\ref{eq:heff}) as 
\begin{align}
&R^{x}(k_{x},\varepsilon)
=
\Xi_{x}(k_{x})
-
\frac{1}{2}
\frac
{(\Delta_{0}+\Delta_{x}(k_{x}))^{2}-(\mathrm{Im}\Delta_{y}(k_{x}))^{2}}
{\varepsilon+\Delta_{0}+\Xi_{x}(k_{x})}
\nonumber \\
&\hspace{43pt}
+
\frac{1}{2}
\frac
{(\Delta_{0}-\Delta_{x}(k_{x}))^{2}-(\mathrm{Im}\Delta_{y}(k_{x}))^{2}}
{\varepsilon+\Delta_{0}-\Xi_{x}(k_{x})}
,
\\
&R^{y}(k_{x},\varepsilon)=0
,
\\
&R^{z}(k_{x},\varepsilon)
=
\frac
{2\{\Delta_{0}\Xi_{x}(k_{x})-\Delta_{x}(k_{x})(\varepsilon+\Xi_{0})\}}
{(\varepsilon+\Xi_{0})^{2}-(\Xi_{x}(k_{x}))^{2}}
\mathrm{Im}\Delta_{y}(k_{x})
\label{eq:rz}
.
\end{align}
We first find that for a collinear antiferromagnet ($\zeta=0$),
$R^{x}(k_{x},\varepsilon)=0$ is realized since $\Xi_{x}(k_{x})=\Delta_{0}=0$.
As another important point, we see from Eqs.(\ref{eq:dy}) and (\ref{eq:rz})
that $R^{z}(k_{x},\varepsilon) \propto D$. 
Therefore, referring to Fig. \ref{fig1}(e), one can conclude that $D\neq0$ 
and a noncollinear magnetic order generate type-(iii) magnon spin textures.
The vector $\bm{R}(k_{x},\varepsilon)$ also provides us the information on the
$k_{x}$-dependence of the spin texture:
$R^{z}(k_{x},\varepsilon)$ is odd with respect to $k_{x}$, indicating that
the $z$-component of the spin is reversed at $k_{x}=0$.
\par
Figure \ref{fig2}(b) shows the magnon bands and spin textures, 
which are explained well from the classification in Fig.\ref{fig1}(e).
This model is a prototype platform, where 
depending on the collinear/noncollinear alignment or $D$-values, all types of textures
are systematically realized.
%
\begin{figure*}[tbp]
\includegraphics[width=180mm]{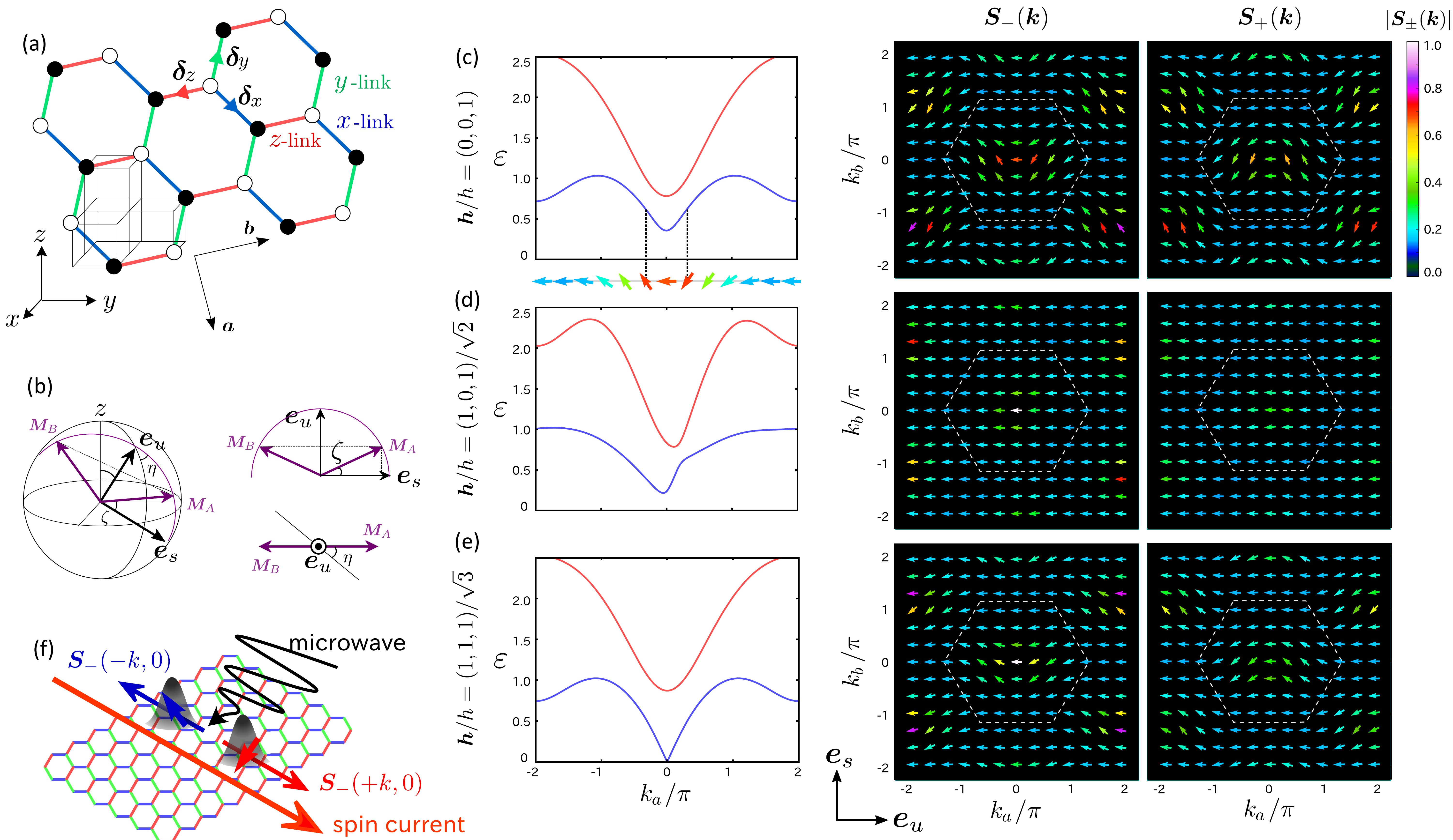}
\caption{(a) Kitaev-Heisenberg-$\Gamma$ model, with its spin axis taken along the $x,y,z$-coordinate. 
The spatial coordinate of the 2D plane is defined along the $a,b$-axis. 
(b) Relation between $\bm{e}_{u}$, $\bm{e}_{s}$ and $\bm{M}_{A/B}$, and the schematic illustration of the
magnetic moments which point in the direction that minimizes the ground state energy.
(c-e) Magnon bands along the $k_{y}=0$ line and spin carried by magnons $\bm{S}_{\pm}(\bm{k})$,
for field angle, (c) $\bm{h}/h=(0,0,1)$, (d) $\bm{h}/h=(1,0,1)/\sqrt{2}$ and (e) $(1,1,1)/\sqrt{3}$, with $h=0.5$.
We set other parameters as $\Theta=\pi/3$ and $\Gamma=-0.1$.
The direction of spins is depicted by taking $\bm{e}_{u}$ and $\bm{e}_{s}$ along the directions shown together, 
and the hexagons in broken lines give the Brillouin zone. 
(f) Schematic illustration of the setup of generating a magnon spin current.
A microwave can excite magnons with a finite wave number $\pm k_{x}$,
whose spins are nearly antiparallel and contribute to a pure spin current.
}
\label{fig3}
\end{figure*}
\begin{figure}[tbp]
\includegraphics[width=90mm]{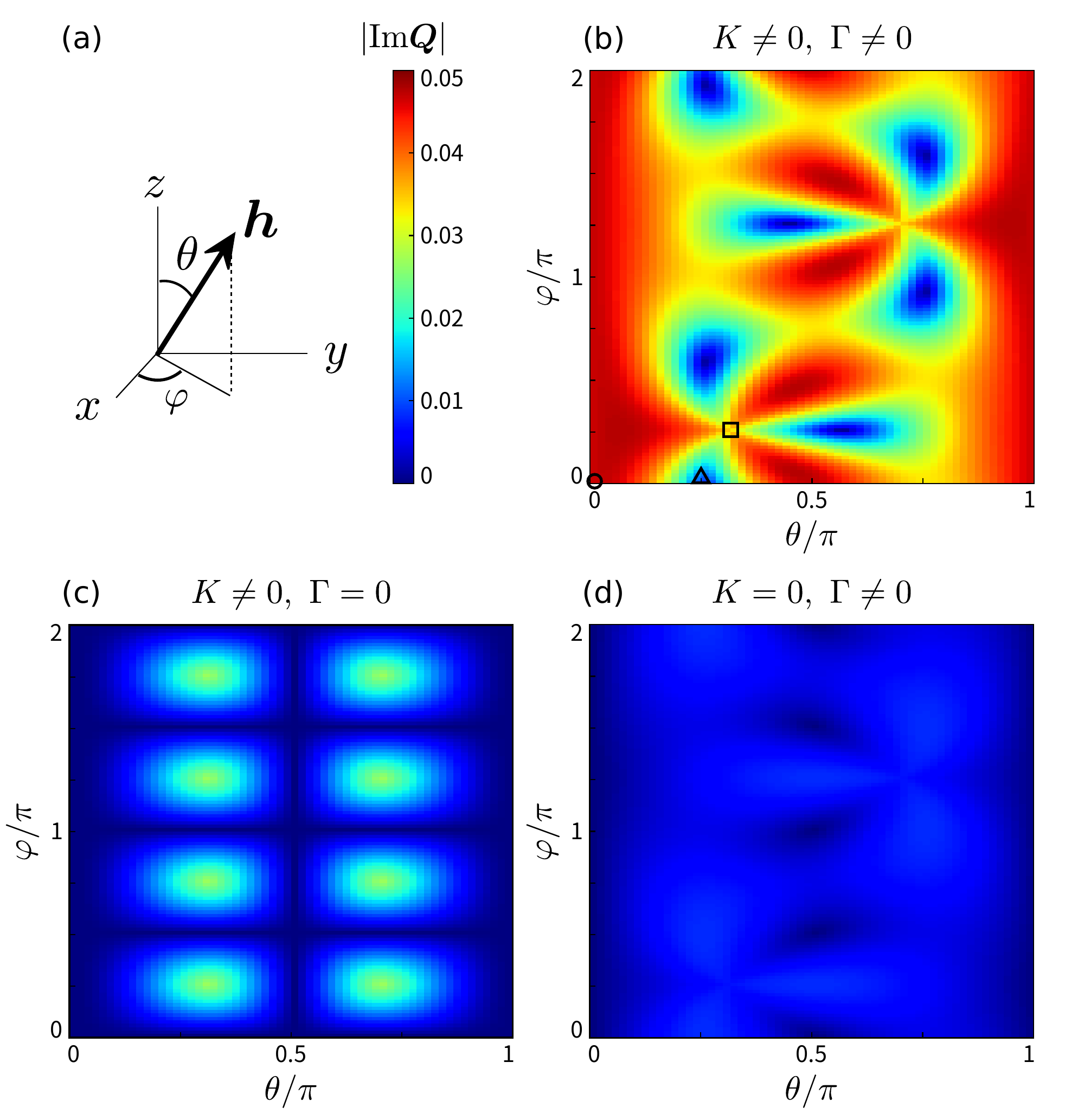}
\caption{(a) Direction of the magnetic field.
(b-d) Density plot of $|\mathrm{Im}\bm{Q}|$ as the function of field angles, $\theta$ and $\varphi$
for (b) $K\neq0$, $\Gamma\neq0$, (c) $K\neq0$, $\Gamma=0$ and (d) $K=0$, $\Gamma\neq0$.
The circle, triangle and rectangle in (b) correspond to $\bm{h}/h=(0,0,1)$, $\bm{h}/h=(1,0,1)/\sqrt{2}$
and $\bm{h}/h=(1,1,1)/\sqrt{3}$, respectively.
}
\label{fig4}
\end{figure}
\subsection{Kitaev-Heisenberg-$\Gamma$ model}
\subsubsection{Kitaev magnons}
The Hamiltonian of the $S=1/2$ KH$\Gamma$ model on the honeycomb lattice 
in a magnetic field is given as \cite{jackeli2009,chaloupka2010,rau2014}
\begin{align}
\hat{\mathcal{H}}
&=
\hspace{-5pt}
\sum_{\mu=x,y,z}
\sum_{\braket{i,j}_{\mu}}
\left\{
J
\hat{\bm{S}}_{i}
\cdot
\hat{\bm{S}}_{j}
+
2K
\hat{S}_{i}^{\mu}
\hat{S}_{j}^{\mu}
\right.
\nonumber \\
&\hspace{50pt}
\left.
+
\Gamma
\left(
\hat{S}_{i}^{\nu}
\hat{S}_{j}^{\rho}
+
\hat{S}_{i}^{\nu}
\hat{S}_{j}^{\rho}
\right)
\right\}
-
\sum_{i}
\bm{h}
\cdot
\hat{\bm{S}}_{i}
,
\label{eq:hamkhg}
\end{align}
where $K$ denotes the Kitaev-type exchange interaction, 
which couples only the $\mu$-component of neighboring spin operators 
along the bonds, $\langle i,j\rangle_\mu$ denoted as $\mu$-link in Fig. \ref{fig3}(a).
The $\Gamma$-term couples the other two spin components,
$(\nu,\rho) \ne \mu$, and $\nu\neq \rho$ along the same $\mu$-link,
and we focus on $\Gamma<0$, which corresponds to Li$_{2}$PrO$_{3}$ \cite{jang2019}.
We parametrize $J$ and $K$ as $J=\cos\Theta$ and $K=\sin\Theta$. 
In the following, we focus on the region, $0\leq\Theta\leq\pi/2$,
where the N\'eel order is realized as a classical ground state
for sufficiently small $\Gamma$ \cite{rau2014,janssen2016}.
The finite magnetic field $\bm{h}$ induces the canting of the antiferromagnetically-ordered moments,
and we denote the canting angle as $\zeta$.
\par
Since the model is highly anisotropic in real space,
the spin axis and the spatial axis are no longer independent with each other.
We adopt the conventional definition for A$_{2}$BO$_{3}$-type layered compounds \cite{jackeli2009},
taking the $x,y,z$-spin axes along the edges of the cube shown in Fig.~\ref{fig3}(a).
These axes are each perpendicular to the $x,y,z$-links of the honeycomb bonds, respectively. 
The two dimensional $a,b$-coordinate of the honeycomb lattice is taken as 
$\bm{a}=(\bm{e}_{x}+\bm{e}_{y}-2\bm{e}_{z})/\sqrt{6}$ and $\bm{b}=(-\bm{e}_{x}+\bm{e}_{y})/\sqrt{2}$ on the honeycomb plane.
\par
The direction of the classical magnetic moments are given as 
\begin{equation}
\bm{M}_{A}=\bm{e}_{u}\sin\zeta+\bm{e}_{s}\cos\zeta
,
\hspace{10pt}
\bm{M}_{B}=\bm{e}_{u}\sin\zeta-\bm{e}_{s}\cos\zeta, 
\end{equation}
where $\bm{e}_{u}$ and $\bm{e}_{s}$ are
the unit vectors pointing in the direction of the uniform and staggered magnetization, respectively. 
The classical energy is given as a function of $\zeta$, $\bm{e}_{u}$ and $\bm{e}_{s}$: 
\begin{align}
E_{\mathrm{cgs}}
&=
-(3J+2K)N_{c}S^{2}\cos2\zeta-2(\bm{h}\cdot\bm{e}_{u})N_{c}S\sin\zeta
\nonumber \\
&\hspace{20pt}
+
2\Gamma N_{c}S^2\sum_{\mu}\left(e_{u}^{\nu}e_{u}^{\rho}\sin^{2}\zeta-e_{s}^{\nu}e_{s}^{\rho}\cos^{2}\zeta\right), 
\label{eq:cgs}
\end{align}
and the ground-state spin configuration is determined by minimizing Eq.(\ref{eq:cgs}).
The canting angle $\zeta$ is then given by
\begin{equation}
\zeta=
\arcsin\left(
\frac
{\bm{h}\cdot\bm{e}_{u}}
{2(3J+2K)S+2\Gamma S\sum_{\mu}(e_{u}^{\nu}e_{u}^{\rho}+e_{s}^{\nu}e_{s}^{\rho})}
\right), 
\end{equation}
and $\bm{e}_{u}$ and $\bm{e}_{s}$ are determined numerically,
where we find that $\bm{e}_{u}$ is almost parallel to $\bm{h}$.
When the magnetic field is perpendicular to the honeycomb plane, $\bm{h}/h=(1,1,1)/\sqrt{3}$,
or $\Gamma=0$, the classical ground state is degenerate with respect to the rotation of $\bm{e}_{s}$ about the $\bm{e}_{u}$-axis.
In this case, one needs to include the zero-point fluctuation energy
to determine the direction of the magnetic moments in the ground state.
The details will be given in Appendix \ref{sec:app_khg}.
\par
The elements of the bosonic BdG Hamiltonian is written as
\begin{align}
&\Xi(\bm{k})=
\Xi_{0}\sigma^{0}+\Xi_{x}(\bm{k})\sigma^{x}+\Xi_{y}(\bm{k})\sigma^{y}+\Xi_{z}
\\
&\Delta(\bm{k})=
\Delta_{x}(\bm{k})\sigma^{x}+\Delta_{y}(\bm{k})\sigma^{y},
\end{align}
with
\begin{align}
&\Xi_{0}=
(3J+2K)S\cos2\zeta
+\bm{h}\cdot\bm{e}_{u}\sin\zeta
\nonumber \\
&\hspace{20pt}
-2\Gamma S\sum_{\mu}\left(e_{u}^{\nu}e_{u}^{\rho}\sin^{2}\zeta-e_{s}^{\nu}e_{s}^{\rho}\cos^{2}\zeta\right)
,
\\
&\Xi_{x}(\bm{k})=
\sum_{\mu}\mathrm{Re}\left[P_{\mu}\mathrm{e}^{i\bm{k}\cdot\bm{\delta}_{\mu}}\right]
,
\label{eq:Xi_x_KHGamma}
\\
&\Xi_{y}(\bm{k})=
-\sum_{\mu}\mathrm{Im}\left[P_{\mu}\mathrm{e}^{i\bm{k}\cdot\bm{\delta}_{\mu}}\right]
,
\label{eq:Xi_y_KHGamma}
\\
&\Xi_{z}=
\bm{h}\cdot\bm{e}_{s}\cos\zeta
,
\label{eq:Xi_z_KHGamma}
\\
&\Delta_{x}(\bm{k})=
\sum_{\mu}Q_{\mu}\cos\bm{k}\cdot\bm{\delta}_{\mu}
,
\label{eq:Delta_x_KHGamma}
\\
&\Delta_{y}(\bm{k})=
-\sum_{\mu}Q_{\mu}\sin\bm{k}\cdot\bm{\delta}_{\mu}
,
\label{eq:Delta_y_KHGamma}
\end{align}
where $\bm{\delta}_{\mu}$ denotes the vector connecting the $\mu$-link (see Fig. \ref{fig3}(a)).
The explicit form of $P_{\mu}$ and $Q_{\mu}$ ($\mu=x,y,z$) are shown in Appendix \ref{sec:app_pq}.
\subsubsection{``Zeeman" field}
From Eq.(\ref{eq:heff}) the vector $\bm{R}(\bm{k},\varepsilon)$ is obtained as 
\begin{align}
R^{x}(\bm{k},\varepsilon)&=
\Xi_{x}(\bm{k})
+\Xi_{x}(-\bm{k})\frac{|\Delta_x(\bm k)|^2-|\Delta_y(\bm k)|^2}{\mathrm{det}[\varepsilon\sigma^0+\Xi^*(-\bm k)]}
\nonumber\\
&\hspace{20pt}
-2\Xi_{y}(-\bm{k})\frac{{\rm Re}(\Delta_x(\bm k)\Delta_y(\bm k))}{\mathrm{det}[\varepsilon\sigma^0+\Xi^*(-\bm k)]}
\label{eq:Rx-kitaev}
\\
R^{y}(\bm{k},\varepsilon)&=
\Xi_y(\bm k)
+2\Xi_x(-\bm k)\frac{{\rm Re}(\Delta_x(\bm k)\Delta_y(\bm k))}{\mathrm{det}[\varepsilon\sigma^0+\Xi^*(-\bm k)]}
\nonumber\\
&\hspace{20pt}
+\Xi_y(-\bm k)\frac{|\Delta_x(\bm k)|^2-|\Delta_y(\bm k)|^2}{\mathrm{det}[\varepsilon\sigma^0+\Xi^*(-\bm k)]}
,
\label{eq:Ry-kitaev}
\\
R^{z}(\bm{k},\varepsilon)&=
\left\{
1-
\frac
{\det\Delta(\bm{k})}
{\det[\varepsilon\sigma^{0}+\Xi^{*}(-\bm{k})]}
\right\}
\Xi_{z}
\nonumber \\
&\hspace{20pt}
-
\frac
{2(\varepsilon+\Xi_{0})\mathrm{Im}[\Delta_{x}^{*}(\bm{k})\Delta_{y}(\bm{k})]}
{\det[\varepsilon\sigma^{0}+\Xi^{*}(-\bm{k})]}
.
\label{eq:Rz-kitaev}
\end{align}
It can be seen that $R^{x}(\bm{k},\varepsilon)$, $R^{y}(\bm{k},\varepsilon)\neq0$ (except at some accidental values) and depend on $\bm k$. 
The first term of $R^{z}(\bm{k},\varepsilon)$ is negligible since 
$\bm{e}_s$ is almost perpendicular to $\bm{h}$, which gives $\Xi^z \simeq 0$.
In the second term, $(\varepsilon+\Xi_{0})$ remains finite in general, and 
\begin{align}
&\mathrm{Im}[\Delta_{x}^{*}(\bm{k})\Delta_{y}(\bm{k})] 
 \nonumber\\
&=\left[\mathrm{Re}\bm{Q}\times\mathrm{Im}\bm{Q}\right]
\cdot
\left[
\begin{pmatrix}
\cos\bm{k}\cdot\bm{\delta}_{x}\\
\cos\bm{k}\cdot\bm{\delta}_{y}\\
\cos\bm{k}\cdot\bm{\delta}_{z}
\end{pmatrix}
\times
\begin{pmatrix}
\sin\bm{k}\cdot\bm{\delta}_{x}\\
\sin\bm{k}\cdot\bm{\delta}_{y}\\
\sin\bm{k}\cdot\bm{\delta}_{z}
\end{pmatrix}
\right]
,
\label{eq:condition_Q}
\end{align}
where $\bm{Q}=(Q_{x},Q_{y},Q_{z})^{T}$. 
Therefore, the necessary and sufficient condition to have $R^{z}(\bm{k},\varepsilon) \ne 0$, 
namely the type-(iii) magnon spin texture, is $\left[\mathrm{Re}\bm{Q}\times\mathrm{Im}\bm{Q}\right]\neq0$. 
For later convenience, we show the form of $\mathrm{Im}\bm{Q}$,
\begin{align}
\mathrm{Im}\bm{Q}
&=
-KS\sin\zeta\ \mathrm{Im}\bm{Q}_{K} - \Gamma S\sin\zeta\ \mathrm{Im}\bm{Q}_{\Gamma}
\nonumber\\ 
\mathrm{Im}\bm{Q}_{K}
&=
\begin{pmatrix}
e_s^x(\bm e_s\times \bm e_u)^x \\
e_s^y(\bm e_s\times \bm e_u)^y \\
e_s^z(\bm e_s\times \bm e_u)^z
\end{pmatrix}
,
\nonumber \\
\mathrm{Im}\bm{Q}_{\Gamma}
&=
\begin{pmatrix}
e_s^y(\bm e_s\times \bm e_u)^z+e_s^z(\bm e_s\times \bm e_u)^y \\
e_s^z(\bm e_s\times \bm e_u)^x+e_s^x(\bm e_s\times \bm e_u)^z \\
e_s^x(\bm e_s\times \bm e_u)^y+e_s^y(\bm e_s\times \bm e_u)^x
\end{pmatrix}. 
\label{eq:imq}
\end{align}
Note that $\mathrm{Im}\bm{Q}_{K}$ and $\mathrm{Im}\bm{Q}_{\Gamma}$ only depend on $\bm{e}_{u}$ and $\bm{e}_{s}$,
which are controlled by the direction of the magnetic field,
and $\mathrm{Im}\bm{Q}$ is scaled by $K$, $\Gamma$, and $\zeta$.
\par
We first consider the Heisenberg limit, $K=\Gamma=0$. 
In this case $\bm Q= -JS\cos^2\zeta(1,1,1)^{T}$ and we find $\mathrm{Im}\bm{Q}=0$, 
which leads to $R^{z}(\bm{k},\varepsilon)=0$ and the type-(ii) magnon spin texture emerges.
When $K\neq0$ and $\Gamma\neq0$,
a finite $J$ still gives $\mathrm{Re}\bm{Q}\neq0$ (see Appendix \ref{sec:app_pq}),
and it can be immediately seen from Eq.(\ref{eq:imq}) that
$\mathrm{Im}\bm{Q}\neq\bm{0}$ holds for any $\bm{e}_{u}$ and $\bm{e}_{s}$.
Namely, $R^{z}(\bm{k},\varepsilon)$ remains nonzero and varies with $\bm{k}$,
which leads to type-(iii) magnon spin texture. 
\par
Let us see the implication of these judgments based on $\bm{R}(\bm{k},\varepsilon)$.
Figure \ref{fig3}(c)-\ref{fig3}(e) shows the spin textures of the KH$\Gamma$ model when $\Theta=\pi/3$ ($K\neq0$) and $\Gamma=-0.1$.
We choose three different field directions, 
$\bm{h}/h=(0,0,1)$, $(1,0,1)/\sqrt{2}$ and $(1,1,1)/\sqrt{3}$.
In the third case, $\bm{h}$ is perpendicular to the honeycomb plane.
One distinct feature of this model is that the spin textures change quite drastically near $\bm{k}=0$ as one can see in Fig. \ref{fig3}(c),
which is not found in the Rashba- and Dresselhaus-type antiferromagnetic insulators on a square lattice \cite{kawano2019-1}.
From Eq.(\ref{eq:condition_Q}) one finds that at around $\bm k\sim 0$, 
$\mathrm{Im}[\Delta_{x}^{*}(\bm{k})\Delta_{y}(\bm{k})]$ changes its sign,
which contributes to the directional variation of $\bm S_\pm(\bm k\sim 0)$
if $\mathrm{Re}\bm{Q}\times\mathrm{Im}\bm{Q}$ is large enough.
The origin of the type-(iii) magnon spin textures is the bond-dependent anisotropic exchange interactions,
and for $\bm{k}\sim0$ the three different directions equally contribute to
$\bm{S}_{\pm}(\bm{k})$, which explains this particular behavior.
\par
The advantage of such large variation of $\bm{S}_{\pm}(\bm{k})$ near $\bm k\sim 0$ is that
most of the experimental tools that excite the magnons can deal only up to some limited energy scale that corresponds to the 
wavelength near $\bm k\sim 0$.
When the magnons at $\bm{k}$ and $-\bm k$ with the same energy are excited as shown in Fig. \ref{fig3}(f),
they carry the spins pointing in the nearly opposite directions, and propagate in the opposite
directions in space.
Since the magnons are charge-neutral and well-defined quasiparticles,
this will generate a dissipationless pure spin current \cite{kawano2019-1}.
The Kitaev materials can thus be an ideal platform of spintronics. 
\par
Another notable advantage of small-$\bm{k}$ magnons is the absence of the damping effect induced by magnon-magnon interactions.
The KH$\Gamma$ model does not have the spin-rotational symmetry, which gives rise to three-magnon interactions \cite{zhitomirsky2013}.
Particularly, the decay process yields the imaginary part of the self-energy
and magnons are strongly damped when the energy and momentum conservation is satisfied.
Since for large $\bm{k}$, or namely at higher-energy part of the bands,
there exists the decay channel that satisfies the energy and momentum conservation,
and the magnons are subject to the strong damping effect.
For the present choices of $\pm\bm{k}$, such effect is avoided. 
\par
To maximize the variation of $R^{z}(\bm{k},\varepsilon)$,
one at least needs to have large enough values of $\mathrm{Im}\bm{Q}$ in Eq.(\ref{eq:condition_Q}).
Figure \ref{fig4}(b)-\ref{fig4}(d) shows the density plot of $|\mathrm{Im}\bm{Q}|$.
This quantity depends on the field direction (while the amplitude is scaled by $h$ regardless of $\bm{k}$).
At the points where $|\mathrm{Im}\bm{Q}|$ takes the maximum, the spin textures show large variation
as shown in Fig.\ref{fig3}(c), whereas when $|\mathrm{Im}\bm{Q}|$ is small,
the spin texture looks almost similar to type-(ii) as shown in Fig.\ref{fig3}(d)
even though there is a slight variation in the direction of $\bm S_\pm(\bm k)$.
By controlling the field direction, one can thus design a desirable spin texture in the KH$\Gamma$ model,
which is a useful reference for experimental studies. 
\par
Notice that the cases, $K\neq0$ and $\Gamma=0$, do not necessarily lead to type-(iii).
Particularly, when the staggered magnetic moment points in the direction perpendicular to
either of the $x,y,z$-links, the propagation of the magnons along that link is suppressed, giving type-(ii).
The details are given in Appendix. \ref{sec:app_khg}.
%
%
\begin{figure*}[tbp]
\includegraphics[width=180mm]{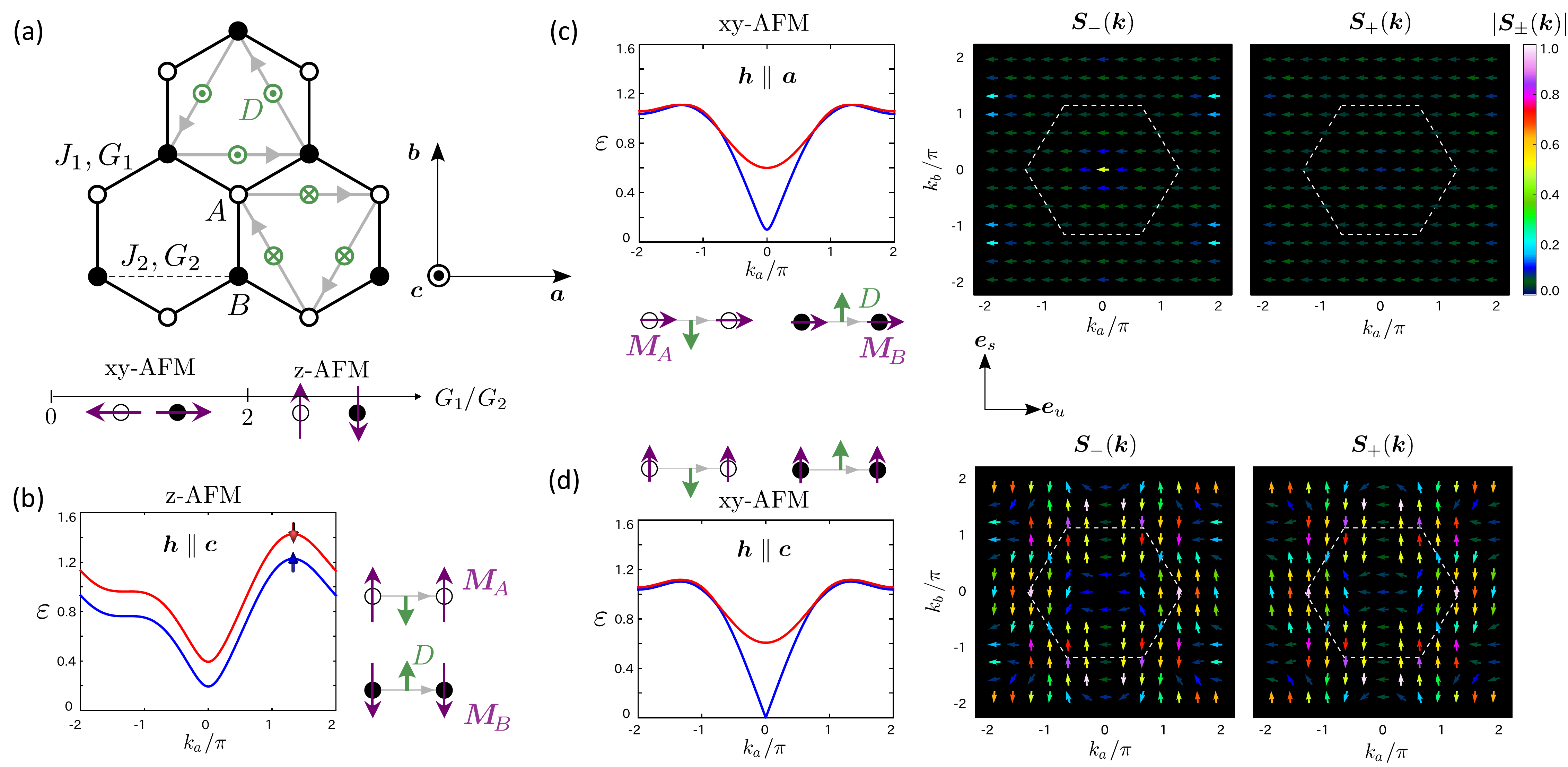}
\caption{(a) Honeycomb-lattice antiferromagnet in Eq.(\ref{eq:ham_honey}). 
Green and gray arrows indicate the direction of the DM vector $\bm D_{i,j}$ 
and the direction $i\rightarrow j$, respectively. 
We set the parameters as $S=1/2$, $J_{1}=1.0$, $J_{2}=0.1$, $G_{2}=0.02$, $h=0.1$
and $G_{1}=0.05$ for the z-AFM phase, and $G_{1}=0$ for the xy-AFM phase. 
The bottom panel is the phase diagram of the model at $h=0$. 
(b) Magnon bands of z-AFM phase along the $k_{a}$-direction ($k_{b}=0$)
and the relative relationships between $\bm{M}_{A}, \bm{M}_{B}$ and $\bm{D}_{i,j}$. 
(c-d) Magnon bands in the xy-AFM phase and the corresponding spin textures $\bm{S}_{\pm}(\bm{k})$. 
For $\bm{h}\parallel \bm{a}$ we have type-(ii) and for $\bm{h}\parallel \bm{c}$ we have type-(iii) magnon spin textures.}
\label{fig5}
\end{figure*}

\subsection{honeycomb lattice antiferromagnet}
\label{sec:honeycomb}
We finally consider an antiferromagnet on the honeycomb lattice
introduced in Ref. \cite{hayami2016}, 
whose Hamiltonian is given as
\begin{align}
\hat{\mathcal{H}}
&=
\sum_{i<j}
J_{i,j}
\hat{\bm{S}}_{i}
\cdot
\hat{\bm{S}}_{j}
+
\sum_{i<j}
G_{i,j}
(
-
\hat{S}_{i}^{x}\hat{S}_{j}^{x}
-
\hat{S}_{i}^{y}\hat{S}_{j}^{y}
+
\hat{S}_{i}^{z}\hat{S}_{j}^{z}
)
\nonumber \\
&\hspace{10pt}
+
\sum_{i<j}
\bm D_{i,j}
\cdot
(\hat{\bm{S}}_{i}\times\hat{\bm{S}}_{j})
-
\sum_{i}
\bm{h}
\cdot
\hat{\bm{S}}_{i}
,
\label{eq:ham_honey}
\end{align}
where we consider the nearest-neighbor and next-nearest-neighbor interactions for both $J_{i,j}$ and $G_{i,j}$,
which are denoted as $J_1,J_2$ and $G_1,G_2$.
The DM vectors $\bm D_{i,j}$ are perpendicular to the honeycomb plane and are 
depicted in Fig. \ref{fig5}(a) by taking the direction of $i\rightarrow j$ as arrows between the next nearest neighbors. 
This Hamiltonian originates from the Hubbard model with SOC
as a result of perturbation from the strong coupling limit \cite{hayami2016}.
The parameters $G_2$ and $D=|\bm{D}_{i,j}|$ come from the SOC, which are related to each other as $D=2\sqrt{J_2G_2}$. 
Here, the DM interaction takes the role of the SOC in the Kane-Mele model \cite{kane2005, kim2016, owerre2016},
namely the propagation of magnons along the next-nearest neighbor bonds couples to
an effective U(1) gauge field, and a staggered flux is inserted between the $A$ and $B$ sublattices.
This staggered flux is the origin of $R^{z}(\bm{k},\varepsilon)\neq0$.
However, the problem is not as simple;
to generate such flux, the relative angle of $\bm{D}_{i,j}$ and $\bm{M}_{A}, \bm{M}_{B}$ should be properly set,
as we see in the following.
\par
The phase diagram of the classical ground state for $\bm{h}=0$ and for sufficiently small $G_{1}$ and $G_{2}$
is shown in the bottom panel of Fig. \ref{fig5}(a);
the collinear antiferromagnetic order is realized, whose moments point in either the $xy$-plane and $z$-directions
for $G_1$ smaller and larger than $2G_2$, respectively, which are called xy-AFM and z-AFM phase \cite{hayami2016}.
\par
We first consider the z-AFM phase with the out-of-plane magnetic field $\bm{h}/h=(0,0,1)$.
For sufficiently small $h$, the ground-state spins remain collinear,
and the elements of the bosonic BdG Hamiltonian is given in the form, 
\begin{align}
&\Xi(\bm{k})=
\Xi_{0}(\bm{k})\sigma^{0}+h\sigma^{z}
\\
&\Delta(\bm{k})=
\Delta_{x}(\bm{k})\sigma^{x}+\Delta_{y}(\bm{k})\sigma^{y}
.
\label{eq:bdg_honey}
\end{align}
The details of $\Xi_{0}(\bm{k})$, $\Delta(\bm{k})$ are shown in Appendix \ref{app:honeycomb}.
This form immedeiately leads to $R^{x}(\bm{k},\varepsilon)=R^{y}(\bm{k},\varepsilon)=0$ and
\begin{equation}
R^{z}(\bm{k},\varepsilon)=
\left\{
1-\frac
{\det\Delta(\bm{k})}
{\det[\varepsilon+\Xi(-\bm{k})]}
\right\}h
.
\end{equation}
From Fig. \ref{fig1}(e), the collinear antiferromagnet with such types of $\bm{R}(\bm{k},\varepsilon)$ gives 
type-(i) magnon spin texture, namely a quantized spins $\bm{S}_{\pm}(\bm{k})=\mp\bm{e}_{z}$.
The energy bands split uniformly, 
$\varepsilon_{\pm}(\bm{k})=\pm h+\sqrt{(\Xi_{0}(\bm{k}))^{2}-(\Delta_{x}(\bm{k}))^{2}-(\Delta_{y}(\bm{k}))^{2}}$, 
as a result of typical Zeeman splitting (see Fig. \ref{fig5}(c)). 
\par
Next, we consider the xy-AFM phase. 
The elements of the bosonic BdG Hamiltonian is given by
\begin{align}
\Xi(\bm{k})
&=
\Xi_{0}(\bm{k})\sigma^{0}+\Xi_{x}(\bm{k})\sigma^{x}+\Xi_{y}(\bm{k})\sigma^{y}+\Xi_{z}(\bm{k})\sigma^{z}
\\
\Delta(\bm{k})
&=
\Delta_{0}(\bm{k})\sigma^{0}+\Delta_{x}(\bm{k})\sigma^{x}+\Delta_{y}(\bm{k})\sigma^{y}. 
\label{eq:bdg_honey}
\end{align}
When the in-plane magnetic field is applied in the $\bm a$-direction, 
the canting angle is calculated as
$\zeta=\arcsin\big(h/(6(J_{1}-G_{1})S))$, and we find $\Xi_{z}(\bm k)=0$. 
The other parameters, $\Xi_{0}(\bm{k}), \Xi_{x}(\bm{k}), \Xi_{y}(\bm{k})$ and 
$\Delta_{0}(\bm{k})$, $\Delta_{x}(\bm{k})$, $\Delta_{y}(\bm{k})$ do not depend on $\bm{D}_{i,j}$
(see Appendix \ref{app:honeycomb}, Eqs.(\ref{eq:honeycomb-xi0})-(\ref{eq:honeycomb-deltay}) for details). 
It is known that the effective U(1) gauge field emerges when $\bm{D}_{i,j}$ has an element parallel to the ordered moments
\cite{onose2010,kawano2019-2}. 
This does not apply to the present $\bm M_A$ and $\bm M_B$, and resultantly 
the magnon excitation is not affected by the DM interaction. 
The ``Zeeman field'' $\bm{R}(\bm{k},\varepsilon)$ is given as
\begin{align}
R^{x}(\bm{k},\varepsilon)
&=
\Xi_{x}(\bm{k})
-\frac{2(\varepsilon+\Xi_{0}(\bm{k}))\Delta_{0}(\bm{k})\Delta_{x}(\bm{k})}{\mathrm{det}[\varepsilon\sigma^{0}+\Xi(\bm{k})]}
\nonumber \\
&\hspace{20pt}
-\Xi_{x}(\bm{k})\frac{(\Delta_{0}(\bm{k}))^{2}+(\Delta_{x}(\bm{k}))^{2}-(\Delta_{y}(\bm{k}))^{2}}{\mathrm{det}[\varepsilon\sigma^{0}+\Xi(\bm{k})]}
\nonumber \\
&\hspace{20pt}
-2\Xi_{y}(\bm{k})\frac{\Delta_{x}(\bm{k})\Delta_{y}(\bm{k})}{\mathrm{det}[\varepsilon\sigma^{0}+\Xi(\bm{k})]}
\label{eq:rx-honeyxy}
\\
R^{y}(\bm{k},\varepsilon)&=
\Xi_{y}(\bm{k})
-\frac{2(\varepsilon+\Xi_{0}(\bm{k}))\Delta_{0}(\bm{k})\Delta_{y}(\bm{k})}{\mathrm{det}[\varepsilon\sigma^{0}+\Xi(\bm{k})]}
\nonumber \\
&\hspace{20pt}
-2\Xi_{x}(\bm{k})\frac{\Delta_{x}(\bm{k})\Delta_{y}(\bm{k})}{\mathrm{det}[\varepsilon\sigma^{0}+\Xi(\bm{k})]}
\nonumber \\
&\hspace{20pt}
-\Xi_{y}(\bm{k})\frac{(\Delta_{0}(\bm{k}))^{2}-(\Delta_{x}(\bm{k}))^{2}+(\Delta_{y}(\bm{k}))^{2}}{\mathrm{det}[\varepsilon\sigma^{0}+\Xi(\bm{k})]}
\label{eq:ry-honeyxy}
\\
R^{z}(\bm{k},\varepsilon)&=0.
\end{align}
Then, from Fig. \ref{fig1}(e) this noncollinear antiferromagnetic order with $R^{z}(\bm{k},\varepsilon)=0$ has type-(ii) spin texture,
as we actually find by calculating $\bm S_{\pm}(\bm k)$ in Fig. \ref{fig5}(c).
\par
If we apply the out-of-plane magnetic field in the $\bm c$-direction 
 to the xy-AFM, the spins cant off the plane with the angle 
$\zeta=\arcsin\big(h/(6(J_{1}+2G_{2})S)\big)$. 
The DM interaction starts to affect the magnon excitation since $\bm{M}_{A}$ and $\bm{M}_{B}$ have the element parallel to $\bm{D}_{i,j}$,
and among the parameters of the bosonic BdG Hamiltonian, $\Xi_z(\bm k)$ starts to depend on $D$ as
\begin{equation}
\Xi_{z}(\bm{k})=-2DS\sin\zeta
\mathrm{Im}
\left[
\sum_{\bm{\delta}'}
\mathrm{e}^{i\bm{k}\cdot\bm{\delta}'} 
\right]
,
\end{equation}
where $\bm{\delta}'_{\mu}$ ($\mu=x,y,z$) denotes the vector connecting the next-nearest neighbor bonds (gray arrows in Fig. \ref{fig5}(a)).
Other parameters still do not depend on $D$ as given
in Eqs.(\ref{eq:honeycomb-xi0-2})-(\ref{eq:honeycomb-deltay-2}) in Appendix \ref{app:honeycomb}. 
The form of $R^{x}(\bm{k},\varepsilon)$ and $R^{y}(\bm{k},\varepsilon)$ remains unchanged from Eqs.(\ref{eq:rx-honeyxy}) and (\ref{eq:ry-honeyxy}),
but $R^{z}(\bm{k},\varepsilon)$ starts to vary with $\bm k$ as 
\begin{equation}
R^{z}(\bm{k},\varepsilon)=
\left\{
1-
\frac
{\mathrm{det}\Delta(\bm{k})}
{\det[\varepsilon\sigma^{0}+\Xi(\bm{k})]}
\right\}\Xi_{z}(\bm{k}). 
\end{equation}
which gives type-(iii) magnon spin textures from Fig.\ref{fig1}(e).
Since $\Xi_{z}(\bm{k})\propto D$, one can immediately see that $\bm D$ is responsible 
for this particular magnon spin texture which is shown in Fig. \ref{fig5}(d).
\par
Finally, let us discuss the physical implication of the role of the DM interaction.
In the Kane-Mele model, when the electrons hop along the next-nearest neighbor bonds in Fig.\ref{fig5}(a)
a SOC works as a staggered magnetic field placed on $A$ and $B$ sublattices \cite{kane2005}.
If we apply the same context to our model,
the DM vectors that generate an effective U(1) gauge field can be regarded as a ``magnetic field'' \cite{kim2016, owerre2016},
as long as $\bm D$ has a component parallel to $\bm M_A$ and $\bm M_B$.
If such a field is placed in a staggered manner on $A$ and $B$ sublattices,
it will generate a fictitious SOC to the pseudo-spins of magnons and generate $R^{z}(\bm{k},\varepsilon)\neq0$.
For the z-AFM, the DM vectors along the $A$-$A$ and $B$-$B$ bonds always point in the opposite direction 
to their magnetic moments, $\bm M_A$ and $\bm M_B$, 
as shown in Fig.\ref{fig5}(b), and do not work as a staggered field. 
For the xy-AFM phase, when the magnetic moment points off the plane by putting $\bm h \parallel c$, 
as shown in Fig.\ref{fig5}(d), the staggered $\bm D$ vectors parallel to the small canted moments 
definitely generate a staggered U(1) gauge field and make $R^{z}(\bm{k},\varepsilon)\neq0$.

%
\section{conclusion}
\label{sec:conclusion}
We proposed a general theoretical framework to design the momentum-dependent spin textures 
for any kinds of complicated insulating antiferromagnets,
in a way as easy as dealing with electronic systems with SOC. 
For a given Hamiltonian of insulating magnets, we first apply a spin-wave theory and 
perform a standard calculation to derive the bosonic BdG hamiltonian (Eq.(\ref{eq:H_BdG})) that describes the magnon excitation.
Then, by transforming the bosonic BdG Hamiltonian to the $2\times2$ Hermitian form $H_{\mathrm{eff}}(\bm{k},\varepsilon)$
which is the essential point of the framework, 
one can describe the Hamiltonian in terms of the pseudo-spin degrees of freedom, $\bm{\sigma}$;
the ``up/down" state of pseudo-spin indicates that the magnon state has a full weight on
the A/B-sublattice, which is described by the north/south pole of the Bloch sphere.
Various terms in the original Hamiltonian is transformed to $\bm{R}(\bm{k},\varepsilon)$
which couple to the pseudo-spin $\bm{\sigma}$ as ``Zeeman'' term $\bm{R}(\bm{k},\varepsilon)\cdot\bm{\sigma}$,
and change the direction of the pseudo-spin as well as the total magnon density the state carries.
When the pseudo-spin on the Bloch sphere point along the equator, the $A$ and $B$ sublattices equally contribute to the magnon eigenstates.
Since the $\hat{a}$- and $\hat{b}$-magnons carry spin moment that point in the directions, $-\bm M_A$ and $-\bm M_B$, respectively.
Therefore, how $\bm{R}(\bm{k},\varepsilon)$ behaves as a function of $\bm{k}$ will determine
the magnitude and direction of the spin moments $\bm{S}_{\pm}(\bm{k})$ in momentum space.
We classified $\bm{S}_{\pm}(\bm{k})$ into types-(i), (ii) and (iii), and showed which types will appear
for given types of $\bm R(\bm{k},\varepsilon)$.
\par
This framework enables us to deal with complicated models and clarify
which parameter or which field direction will generate a favorable spin texture,
in a simple enough manner without knowing in advance the role of interactions included in the Hamiltonian. 
This is actually demonstrated in the Kitaev-Heisenberg-$\Gamma$ model. 
We discovered that the Kitaev interaction generates a type-(iii) spin
texture that varies significantly with $\bm{k}$ assisted by the $\Gamma$-term.
This will shed light on the useful aspect of a series of Kitaev materials,
$A_{2}$PrO$_{3}$ ($A$=Li, Na), on the top of the possible realization of the prototype $\mathbb{Z}_{2}$ spin liquid
and Berry-phase effect on magnons \cite{mcclarty2018,joshi2018,cookmeyer2018,choi2019}.
\par
So far, most of the studies on spin textures are performed in electronic systems with SOC.
There has been a growing quest for finding a better and easier platform to control spin moments,
particularly in antiferromagnets, since the nearly net-zero magnetic moment is convenient to
efficiently reduce the effect of a stray field.
The present findings provide a clue to design or test unexplored materials as a candidate for forthcoming spintronics.
\par
\begin{acknowledgments}
The authors thank Yuki Shiomi for helpful comments.
This work is supported by JSPS KAKENHI Grants No. JP17K05533, No. JP18H01173, No.JP17K05497, No. JP17H02916.
M. K was supported by Grant-in-Aid for JSPS Research Fellow (Grant No. 19J22468).
\end{acknowledgments}

\appendix

\section{Bosonic BdG Hamiltonian}
\label{sec:bdg}
We show the details of the spin-wave analysis
for the system with a two-sublattice long-range magnetically-ordered ground state.
The low-energy excitation can be represented by magnons using a Holstein-Primakoff transformation \cite{holstein1940}:
\begin{align}
\hat{\bm{S}}_{i\in A}
&\simeq
(S-\hat{a}_{i}^{\dagger}\hat{a}_{i})
\bm{M}_{A}
\nonumber \\
&\hspace{10pt}
+
\sqrt{\frac{S}{2}}
\left(
\hat{a}_{i}
+
\hat{a}_{i}^{\dagger}
\right)
\bm{X}_{A}
-
i
\sqrt{\frac{S}{2}}
\left(
\hat{a}_{i}
-
\hat{a}_{i}^{\dagger}
\right)
\bm{Y}_{A}
,
\\
\hat{\bm{S}}_{i\in B}
&\simeq
(S-\hat{b}_{i}^{\dagger}\hat{b}_{i})
\bm{M}_{B}
\nonumber \\
&\hspace{10pt}
+
\sqrt{\frac{S}{2}}
\left(
\hat{b}_{i}
+
\hat{b}_{i}^{\dagger}
\right)
\bm{X}_{B}
-
i
\sqrt{\frac{S}{2}}
\left(
\hat{b}_{i}
-
\hat{b}_{i}^{\dagger}
\right)
\bm{Y}_{B}
.
\end{align}
The unit vectors, $\bm{X}_{A/B}$, $\bm{Y}_{A/B}$ and $\bm{M}_{A/B}$,
form the basis of the local right-handed coordinate system on the sublattice $A/B$,
and $\bm{M}_{A/B}$ points in the direction of the classical spin in the ground state. 
Fourier transformations of $\hat{a}_{i}$/$\hat{b}_{i}$ are given by
\begin{equation}
\hat{a}_{i}
=
\frac{1}{\sqrt{N_{c}}}
\sum_{\bm{k}}
\hat{a}(\bm{k})
\mathrm{e}^{i\bm{k}\cdot\bm{r}_{i}}
,
\hspace{10pt}
\hat{b}_{i}
=
\frac{1}{\sqrt{N_{c}}}
\sum_{\bm{k}}
\hat{b}(\bm{k})
\mathrm{e}^{i\bm{k}\cdot\bm{r}_{i}}
,
\end{equation}
where $N_{c}$ is the number of unit cells and $\bm{r}_{i}$ is the position at a site $i$.
This transformation gives Eq.(\ref{eq:ham_sw}). 
Magnon bands $\varepsilon_{\pm}(\bm{k})$ and corresponding eigenvectors $\bm{t}_{\pm}(\bm{k})$
can be obtained from the following eigenvalue equation \cite{colpa1978}:
\begin{equation}
\Sigma^{z}H_{\mathrm{BdG}}(\bm{k})
\bm{t}_{\pm}(\bm{k})
=
\varepsilon_{\pm}(\bm{k})
\bm{t}_{\pm}(\bm{k})
,
\end{equation}
with $\Sigma^{\mu}=\tau^{\mu}\otimes\sigma^{0}$, 
where $\tau^{\mu}$ and $\sigma^{\mu}$ ($\mu=0,x,y,z$) are the unit and Pauli matrices acting on a particle-hole space
and sublattice space, respectively. 
The matrix $\Sigma^{z}$ is required here to keep the bosonic statistics.
$H_{\mathrm{BdG}}(\bm{k})$ generally has the following particle-hole symmetry:

\begin{equation}
\Sigma^{x}\Sigma^{z}H_{\mathrm{BdG}}(\bm{k})\Sigma^{x}=-\Sigma^{z}H_{\mathrm{BdG}}(-\bm{k})
.
\end{equation}
Therefore, the matrix $\Sigma^{z}H_{\mathrm{BdG}}(\bm{k})$
also has the eigenvalues $-\varepsilon_{\pm}(-\bm{k})$
and corresponding eigenvectors $\Sigma^{x}\bm{t}_{\pm}^{*}(-\bm{k})$.
We normalize the eigenvector $\bm{t}_{n}(\bm{k})$ as
$\bm{t}_{n}^{\dagger}(\bm{k})\Sigma^{z}\bm{t}_{m}(\bm{k})=\delta_{n,m}$
($n,m=\pm$)
and constract a matrix $T(\bm{k})$ as
\begin{equation}
T(\bm{k})=
(
\bm{t}_{+}(\bm{k})
,
\bm{t}_{-}(\bm{k})
,
\Sigma^{x}\bm{t}_{+}^{*}(-\bm{k})
,
\Sigma^{x}\bm{t}_{-}^{*}(-\bm{k})
)
.
\end{equation}
Then the matrix $T(\bm{k})$ satisfies the paraunitary condition:
$T^{\dagger}(\bm{k})\Sigma^{z}T(\bm{k})=T(\bm{k})\Sigma^{z}T^{\dagger}(\bm{k})=\Sigma^{z}$,
and $\Sigma^{z}H_{\mathrm{BdG}}(\bm{k})$ can be diagonalized by $T(\bm{k})$ as
\begin{equation}
T^{-1}(\bm{k})
\Sigma^{z}
H_{\mathrm{BdG}}(\bm{k})
T(\bm{k})
=
E(\bm{k})
,
\label{bdgdiag}
\end{equation}
\begin{equation}
E(\bm{k})=\mathrm{diag}(\varepsilon_{+}(\bm{k}),\varepsilon_{-}(\bm{k}),-\varepsilon_{+}(-\bm{k}),-\varepsilon_{-}(-\bm{k}))
.
\label{ekvec}
\end{equation}
A Bogoliubov transformation is given by
\begin{equation}
\hat{\Phi}(\bm{k})=T(\bm{k})\hat{\Psi}(\bm{k})
,
\end{equation}
where $\hat{\Psi}(\bm{k})=(\hat{\gamma}_{+}(\bm{k}),\hat{\gamma}_{-}(\bm{k}),\hat{\gamma}_{+}^{\dagger}(-\bm{k}),\hat{\gamma}_{-}^{\dagger}(-\bm{k}))^{T}$
and $\hat{\gamma}_{\pm}(\bm{k})$ also satisfies bosonic commutation relations due to the paraunitary condition for $T(\bm{k})$.
The diagonalized form of $\hat{\mathcal{H}}_{\mathrm{sw}}$ is 
\begin{equation}
\hat{\mathcal{H}}_{\mathrm{sw}}
=
\sum_{\bm{k}}
\sum_{n=\pm}
\varepsilon_{n}(\bm{k})
\left(
\hat{\gamma}_{n}^{\dagger}(\bm{k})
\hat{\gamma}_{n}(\bm{k})
+
\frac{1}{2}
\right)
.
\end{equation}
We now write the eigenvector $\bm{t}_{\pm}(\bm{k})$ as
\begin{equation}
\bm{t}_{\pm}(\bm{k})
=
\begin{pmatrix}
\bm{u}_{\pm}(\bm{k})\\
\bm{v}_{\pm}(\bm{k})
\end{pmatrix}
=
\begin{pmatrix}
u_{\pm,A}(\bm{k})\\
u_{\pm,B}(\bm{k})\\
v_{\pm,A}(\bm{k})\\
v_{\pm,B}(\bm{k})
\end{pmatrix}
.
\end{equation}

\section{Derivation of an effective Hamiltonian}
\label{subsec:H_eff}
We extend the Brillouin-Wigner formalism for fermions
in Eqs. (\ref{eq:H_orbital})-(\ref{eq:H_eff_electron}) to a bosonic case.
Let us consider the Green function for the non-Hermitian matrix $\Sigma^{z}H_{\mathrm{BdG}}(\bm{k})$:
\begin{equation}
G_{\mathrm{BdG}}(\bm{k},\varepsilon)
=
[\varepsilon\Sigma^{0}-\Sigma^{z}H_{\mathrm{BdG}}(\bm{k})]^{-1}
,
\label{eq:G_BdG}
\end{equation}
where $\varepsilon\in\mathbb{R}$, and the matrix $\Sigma^{z}$ is introduced to keep the boson statistics. 
In the same manner as Eq.(\ref{bdgdiag}), 
$G_{\mathrm{BdG}}(\bm{k},\varepsilon)$ can also be diagonalized by $T(\bm{k})$ as
\begin{equation}
T^{-1}(\bm{k})
G_{\mathrm{BdG}}(\bm{k},\varepsilon)
T(\bm{k})
=
[\varepsilon\Sigma^{0}-E(\bm{k})]^{-1}
.
\label{eq:G_BdG_diagonal}
\end{equation}
where $E(\bm{k})$ is the matrix form of the eigen energy in Eq.(\ref{ekvec}).
Then, the spectral decomposition of $G_{\mathrm{BdG}}(\bm{k},\varepsilon)$ is obtained as,  
\begin{align}
&G_{\mathrm{BdG}}(\bm{k},\varepsilon)
\nonumber \\
&=\sum_{n=\pm}
\left(
\frac
{\bm{t}_{n}(\bm{k})\bm{t}_{n}^{\dagger}(\bm{k})\Sigma^{z}}
{\varepsilon-\varepsilon_{n}(\bm{k})}
-
\frac
{\Sigma^{x}\bm{t}_{n}^{*}(-\bm{k})\bm{t}_{n}^{T}(-\bm{k})\Sigma^{x}\Sigma^{z}}
{\varepsilon+\varepsilon_{n}(-\bm{k})}
\right)
.
\label{eq:G_BdG_spectral}
\end{align}
Here, one can see that
the pole and residue of $G_{\mathrm{BdG}}(\bm{k},\varepsilon)$ give the exact eigenvalue
and projected weight onto the corresponding eigenspace of $\Sigma^{z}H_{\mathrm{BdG}}(\bm{k})$, respectively. 
\par
At the same time, by applying the same idea as Eqs. (\ref{eq:H_orbital})-(\ref{eq:H_eff_electron}) and Ref. \cite{asano2011},
we divide $G_{\mathrm{BdG}}(\bm{k},\varepsilon)$ into $2\times2$ blocks,
$G(\bm{k},\varepsilon)$ and $F(\bm{k},\varepsilon)$, as
\begin{equation}
G_{\mathrm{BdG}}(\bm{k},\varepsilon)
=
\begin{pmatrix}
G(\bm{k},\varepsilon) & F(\bm{k},\varepsilon)\\
-F^{\dagger}(\bm{k},\varepsilon) & -G^{*}(-\bm{k},-\varepsilon)
\end{pmatrix}
,
\label{eq2x2gbdg}
\end{equation}
and focus on the particle subspace, 
\begin{equation}
G(\bm{k},\varepsilon)
=
[\varepsilon\sigma^{0}-H_{\mathrm{eff}}(\bm{k},\varepsilon)]^{-1} 
\label{eq:Gke}
\end{equation}
Then, we obtain the $2\times2$ effective Hamiltonian 
$H_{\mathrm{eff}}(\bm{k},\varepsilon)$ in Eq.(\ref{eq:heff}). 
\par
Notice that in the same context as 
Eqs.(\ref{eq:H_orbital})-(\ref{eq:H_eff_electron}), the solution of the pole of Eq.(\ref{eq:Gke}),
namely $\epsilon\sigma^{0}=H_{\mathrm{eff}}(\bm{k},\varepsilon)$, 
gives {\it exactly} the eigenvalues of $H_{\mathrm{BdG}}(\bm{k})$, as we show in the following;
as for the effective Hamiltonian, we have 
\begin{equation}
H_{\mathrm{eff}}(\bm{k},\varepsilon)\bm{w}_{\pm}(\bm{k},\varepsilon)
=
\lambda_{\pm}(\bm{k},\varepsilon)\bm{w}_{\pm}(\bm{k},\varepsilon)
,
\label{eq:secular-Heff}
\end{equation}
with the eigenvalue $\lambda_{\pm}(\bm{k},\varepsilon)=R^{0}(\bm{k},\varepsilon)\pm|\bm{R}(\bm{k},\varepsilon)|$
and the eigenvector $\bm{w}_{\pm}(\bm{k},\varepsilon)$ which is normalized as 
$\bm{w}_{n}^{\dagger}(\bm{k},\varepsilon)\bm{w}_{m}(\bm{k},\varepsilon)=\delta_{n,m}$
($n,m=\pm$).
Using these values, 
the spectral decomposition of $G(\bm{k},\varepsilon)$ is written as
\begin{equation}
G(\bm{k},\varepsilon)=
\sum_{n=\pm}
\frac
{\bm{w}_{n}(\bm{k},\varepsilon)\bm{w}_{n}^{\dagger}(\bm{k},\varepsilon)}
{\varepsilon-\lambda_{n}(\bm{k},\varepsilon)}
.
\label{eq:G_spectral}
\end{equation}
By combining Eqs. (\ref{eq:G_BdG_spectral}) and (\ref{eq2x2gbdg}), 
the same Green function is written also in the form, 
\begin{equation}
G(\bm{k},\varepsilon)=
\sum_{n=\pm}
\left(
\frac
{\bm{u}_{n}(\bm{k})\bm{u}_{n}^{\dagger}(\bm{k})}
{\varepsilon-\varepsilon_{n}(\bm{k})}
-
\frac
{\{\bm{v}_{n}(-\bm{k})\bm{v}_{n}^{\dagger}(-\bm{k})\}^{*}}
{\varepsilon+\varepsilon_{n}(-\bm{k})}
\right)
,
\label{eq:G_from_G_BdG_spectral}
\end{equation}
and one immediately finds 
its poles at $\varepsilon=\varepsilon_{n}(\bm{k})$ and $-\varepsilon_{n}(-\bm{k})$ ($n=\pm$). 
From Eqs.(\ref{eq:G_spectral}) and (\ref{eq:G_from_G_BdG_spectral}), 
\begin{equation}
\varepsilon=\lambda_{n}(\bm{k},\varepsilon)
,
\label{eq:eps=lambda}
\end{equation}
is expected to have two solutions, 
$\varepsilon=\varepsilon_{n}(\bm{k})$ and $\varepsilon=-\varepsilon_{n}(-\bm{k})$, 
which is proved rigorously (see Appendix \ref{sec:app1}). 
The residue of $G(\bm{k},\varepsilon)$ at $\varepsilon=\varepsilon_{n}(\bm{k})$ and $\varepsilon=-\varepsilon_{n}(-\bm{k})$
gives the relation between the eigenstates of $\Sigma^{z}H_{\mathrm{BdG}}(\bm{k})$ and $H_{\mathrm{eff}}(\bm{k},\varepsilon)$ as
\begin{align}
\bm{u}_{n}(\bm{k})
&=
\mathrm{cosh}\chi_{n}(\bm{k})\bm{w}_{n}(\bm{k},\varepsilon_{n}(\bm{k}))
\label{eq:u_w}
\\
\bm{v}_{n}(\bm{k})
&=
\mathrm{sinh}\chi_{n}(\bm{k})\bm{w}_{n}^{*}(-\bm{k},-\varepsilon_{n}(\bm{k}))
\label{eq:v_w}
\\
\cosh\chi_{n}(\bm{k})
&=
\left(
1-\left.\frac{\partial\lambda_{n}(\bm{k},\varepsilon)}{\partial\varepsilon}\right|_{\varepsilon=\varepsilon_{n}(\bm{k})}
\right)^{-1/2}
,
\label{eq:cosh=lambda}
\end{align}
where $\chi_{n}(\bm{k})\geq0$.
For details, see Appendix \ref{sec:app1}. 
In this way, the exact relationships between the eigenstates $\bm{t}_{n}(\bm{k})=(\bm{u}_{n}(\bm{k}),\bm{v}_{n}(\bm{k}))^{T}$ and 
$\bm{w}_{n}(\bm{k},\varepsilon)$ are explictly given in Eqs. (\ref{eq:u_w}), (\ref{eq:v_w})
and these relations allow us to rewrite the difference between the local spectral weight
of magnons on $A$ and $B$ sublattices as
\begin{align}
d_{\pm,A}(\bm{k})
-
d_{\pm,B}(\bm{k})
&=
\pm\left\{
\cosh^{2}\chi_{\pm}(\bm{k})
\frac{R^{z}(\bm{k},\varepsilon_{\pm}(\bm{k}))}{|\bm{R}(\bm{k},\varepsilon_{\pm}(\bm{k}))|}
\right.
\nonumber \\
&\hspace{20pt}
\left.
+
\sinh^{2}\chi_{\pm}(\bm{k})
\frac{R^{z}(-\bm{k},-\varepsilon_{\pm}(\bm{k}))}{|\bm{R}(-\bm{k},-\varepsilon_{\pm}(\bm{k}))|}
\right\}
.
\label{eq:dA-dB}
\end{align}
Namely, $R^{z}(\bm{k},\varepsilon)$
is directly related to the difference between $d_{\pm,A}(\bm{k})$ and $d_{\pm,B}(\bm{k})$.
In a while, the sum of the local spectral weight $d_{\pm,A}(\bm{k})+d_{\pm,B}(\bm{k})$ only depends on $\chi_{\pm}(\bm{k})$,
and it can be written as
\begin{equation}
d_{\pm,A}(\bm{k})+d_{\pm,B}(\bm{k})
=
\cosh2\chi_{\pm}(\bm{k})
.
\end{equation}
%
\section{Relation between a bosonic BdG Hamiltonian and an effective Hamiltonian}
\label{sec:app1}
We first show that the eigenvalue of an effective Hamiltonian $\lambda_{n}(\bm{k},\varepsilon)$ ($n=\pm)$
is a monotonically increasing function with respect to $\varepsilon$.
From the eigenvalue equation (\ref{eq:secular-Heff}),
we find the following expression of $\lambda_{n}(\bm{k},\varepsilon)$,
\begin{equation}
\lambda_{n}(\bm{k},\varepsilon)=
\bm{w}_{n}^{\dagger}(\bm{k},\varepsilon)H_{\mathrm{eff}}(\bm{k},\varepsilon)\bm{w}_{n}(\bm{k},\varepsilon)
,
\end{equation}
where the normalization condition for $\bm{w}_{m}(\bm{k},\varepsilon)$,
$\bm{w}_{m}^{\dagger}(\bm{k},\varepsilon)\bm{w}_{m'}(\bm{k},\varepsilon)=\delta_{m,m'}$ ($m,m'=\pm$), is used.
Then the derivative of $\lambda_{n}(\bm{k},\varepsilon)$ with respect to $\varepsilon$ is calculated as
\begin{align}
%
\frac{\partial\lambda_{n}}{\partial\varepsilon}&=
\frac{\partial\bm{w}_{n}^{\dagger}}{\partial\varepsilon}
H_{\mathrm{eff}}\bm{w}_{\pm}
+\bm{w}_{n}^{\dagger}\frac{\partial H_{\mathrm{eff}}}{\partial\varepsilon}\bm{w}_{n}
+\bm{w}_{n}^{\dagger}H_{\mathrm{eff}}\frac{\partial\bm{w}_{n}}{\partial\varepsilon}
\nonumber \\
&=
\bm{w}_{n}^{\dagger}\frac{\partial H_{\mathrm{eff}}}{\partial\varepsilon}\bm{w}_{n}
+\lambda_{n}\frac{\partial}{\partial\varepsilon}\left\{\bm{w}_{n}^{\dagger}\bm{w}_{n}\right\}
\nonumber \\
&=
\bm{w}_{n}^{\dagger}\frac{\partial H_{\mathrm{eff}}}{\partial\varepsilon}\bm{w}_{n}
.
\label{eq:der-lambda}
\end{align}
Since the $2\times2$ matrix $\Xi(\bm{k})$ in $H_{\mathrm{BdG}}(\bm{k})$ is Hermitian,
there exists the spectral decomposition of $\Xi(\bm{k})$,
\begin{equation}
\Xi(\bm{k})=
\sum_{i=1,2}\xi_{i}(\bm{k})\bm{z}_{i}(\bm{k})\bm{z}_{i}^{\dagger}(\bm{k})
,
\end{equation}
where $\xi_{i}(\bm{k})\in\mathbb{R}$ and $\bm{z}_{i}(\bm{k})\in\mathbb{C}^{2}$
are the eigenvalue and corresponding eigenvector of $\Xi(\bm{k})$.
Using these values, the effective Hamiltonian can be written in the form,
\begin{align}
H_{\mathrm{eff}}(\bm{k},\varepsilon)
&=
\sum_{i=1,2}
\left\{
\xi_{i}(\bm{k})\bm{z}_{i}(\bm{k})\bm{z}_{i}^{\dagger}(\bm{k})
\right.
\nonumber \\
&\hspace{20pt}
\left.
-
\Delta(\bm{k})
\frac{\bm{z}_{i}^{*}(-\bm{k})\bm{z}_{i}^{T}(-\bm{k})}{\varepsilon+\xi_{i}(-\bm{k})}
\Delta^{\dagger}(\bm{k})
\right\}
,
\end{align}
and its derivative with respect to $\varepsilon$ is given by
\begin{equation}
\frac{\partial H_{\mathrm{eff}}(\bm{k},\varepsilon)}{\partial\varepsilon}=
\sum_{i=1,2}
\Delta(\bm{k})
\frac{\bm{z}_{i}^{*}(-\bm{k})\bm{z}_{i}^{T}(-\bm{k})}{\{\varepsilon+\xi_{i}(-\bm{k})\}^{2}}
\Delta^{\dagger}(\bm{k})
.
\label{eq:der-Heff}
\end{equation}
From Eqs. (\ref{eq:der-lambda}) and (\ref{eq:der-Heff}), we find
\begin{equation}
\frac{\partial\lambda_{n}(\bm{k},\varepsilon)}{\partial\varepsilon}=
\sum_{i=1,2}
\frac{|\bm{z}_{i}^{T}(-\bm{k})\Delta^{\dagger}(\bm{k})\bm{w}_{n}(\bm{k},\varepsilon)|^2}
{\{\varepsilon+\xi_{i}(-\bm{k})\}^{2}}
\geq0
,
\end{equation}
indicating that $\lambda_{n}(\bm{k},\varepsilon)$ increases monotonically with respect to $\varepsilon$. 
Since $\lambda_{+}(\bm{k},\varepsilon_{+}(\bm{k}))\geq\lambda_{+}(\bm{k},\varepsilon_{-}(\bm{k}))$
and $\lambda_{+}(\bm{k},\varepsilon)\geq\lambda_{-}(\bm{k},\varepsilon)$, we find 
\begin{equation}
\lambda_{+}(\bm{k},\varepsilon_{+}(\bm{k}))\geq\lambda_{-}(\bm{k},\varepsilon_{-}(\bm{k})). 
\label{eq:ineq-lambda}
\end{equation}
\par
Now we derive the relation between the eigenvalue/eigenstate of $H_{\mathrm{BdG}}(\bm{k})$ and $H_{\mathrm{eff}}(\bm{k},\varepsilon)$
from Eqs. (\ref{eq:G_spectral}) and (\ref{eq:G_from_G_BdG_spectral}).
As one can see from Eq. (\ref{eq:G_from_G_BdG_spectral}),
$G(\bm{k},\varepsilon)$ has the simple pole
at $\varepsilon=\varepsilon_{n}(\bm{k})$ and $\varepsilon=-\varepsilon_{n}(-\bm{k})$ ($n=\pm$).
We first focus on the residue at $\varepsilon=\varepsilon_{n}(\bm{k})$:
\begin{equation}
\lim_{\varepsilon\to\varepsilon_{n}(\bm{k})}(\varepsilon-\varepsilon_{n}(\bm{k}))G(\bm{k},\varepsilon)
,
\end{equation}
which leads to the following relation:
\begin{align}
\bm{u}_{n}(\bm{k})\bm{u}_{n}^{\dagger}(\bm{k})
=&
\sum_{m=\pm}
\left(
\lim_{\varepsilon\to\varepsilon_{n}(\bm{k})}
\frac
{\varepsilon-\varepsilon_{n}(\bm{k})}
{\varepsilon-\lambda_{m}(\bm{k},\varepsilon)}
\right)
\nonumber\\
&\hspace{20pt}
\times\bm{w}_{m}(\bm{k},\varepsilon_{n}(\bm{k}))
\bm{w}_{m}^{\dagger}(\bm{k},\varepsilon_{n}(\bm{k}))
.
\label{eq:uu=sum_ww}
\end{align}
Because of the normalization condition for $\bm{t}_{n}(\bm{k})$,
the vectors $\bm{u}_{n}(\bm{k})$ and $\bm{v}_{n}(\bm{k})$ must satisfy
\begin{equation}
\bm{u}_{n}^{\dagger}(\bm{k})\bm{u}_{n}(\bm{k})
-
\bm{v}_{n}^{\dagger}(\bm{k})\bm{v}_{n}(\bm{k})
=
1
.
\end{equation}
Therefore, $\bm{u}_{n}(\bm{k})$ and $\bm{v}_{n}(\bm{k})$ can be generally written as
\begin{align}
\bm{u}_{n}(\bm{k})
&=
\cosh\chi_{n}(\bm{k})\tilde{\bm{u}}_{n}(\bm{k})
,
\\
\bm{v}_{n}(\bm{k})
&=
\sinh\chi_{n}(\bm{k})\tilde{\bm{v}}_{n}(\bm{k})
,
\end{align}
where $\chi_{n}(\bm{k})\geq0$.
The vectors $\tilde{\bm{u}}_{n}(\bm{k})$ and $\tilde{\bm{v}}_{n}(\bm{k})$ satisfy the following normalization condition:
\begin{equation}
\tilde{\bm{u}}_{n}^{\dagger}(\bm{k})\tilde{\bm{u}}_{n}(\bm{k})=1
,
\hspace{10pt}
\tilde{\bm{v}}_{n}^{\dagger}(\bm{k})\tilde{\bm{v}}_{n}(\bm{k})=1
.
\end{equation}
Then the Eq. (\ref{eq:uu=sum_ww}) can be rewritten as
\begin{equation}
\tilde{\bm{u}}_{n}(\bm{k})\tilde{\bm{u}}_{n}^{\dagger}(\bm{k})
=
\sum_{m=\pm}
L_{n,m}(\bm{k})
\bm{w}_{m}(\bm{k},\varepsilon_{n}(\bm{k}))
\bm{w}_{m}^{\dagger}(\bm{k},\varepsilon_{n}(\bm{k}))
,
\label{eq:uu=cosh_sum_ww}
\end{equation}
with
\begin{equation}
L_{n,m}(\bm{k})
=
\frac{1}{\cosh^{2}\chi_{n}(\bm{k})}
\lim_{\varepsilon\to\varepsilon_{n}(\bm{k})}
\frac
{\varepsilon-\varepsilon_{n}(\bm{k})}
{\varepsilon-\lambda_{m}(\bm{k},\varepsilon)}
.
\label{eq:L_n}
\end{equation}
From Eq. (\ref{eq:uu=cosh_sum_ww}), we find that $\bm{w}_{m}(\bm{k},\varepsilon_{n}(\bm{k}))$ satisfies the following eigenvalue equation:
\begin{equation}
\tilde{\bm{u}}_{n}(\bm{k})\tilde{\bm{u}}_{n}^{\dagger}(\bm{k})
\bm{w}_{m}(\bm{k},\varepsilon_{n}(\bm{k}))
=
L_{n,m}(\bm{k})
\bm{w}_{m}(\bm{k},\varepsilon_{n}(\bm{k}))
,
\label{eq:uuw=w}
\end{equation}
where we use the normalization condition for $\bm{w}_{m}(\bm{k},\varepsilon)$.
One can see that the rank of the $2\times2$ matrix
$\tilde{\bm{u}}_{n}(\bm{k})\tilde{\bm{u}}_{n}^{\dagger}(\bm{k})$ is equal to 1.
Then the only nonzero eigenvalue and corresponding eigenvector
of $\tilde{\bm{u}}_{n}(\bm{k})\tilde{\bm{u}}_{n}^{\dagger}(\bm{k})$ are given by $1$ and $\tilde{\bm{u}}_{n}(\bm{k})$
since $\tilde{\bm{u}}_{n}^{\dagger}(\bm{k})\tilde{\bm{u}}_{n}(\bm{k})=1$.
From these facts and the inequality described by Eq.(\ref{eq:ineq-lambda}),
the eigenvalue equation (\ref{eq:uuw=w}) leads to the following relations,
\begin{align}
&L_{n,m}(\bm{k})
=\delta_{n,m}
,
\label{eq:L=1}
\\
&\bm{w}_{n}(\bm{k},\varepsilon_{n}(\bm{k}))
=
\tilde{\bm{u}}_{n}(\bm{k})
,
\label{eq:w=u}
\end{align}
for $m=\pm$.
From Eq. (\ref{eq:w=u}),
the relation between $\bm{u}_{n}(\bm{k})$ and $\bm{w}_{n}(\bm{k},\varepsilon)$ can be obtained as Eq. (\ref{eq:u_w})
up to an phase factor, which can be included into $\bm{w}_{n}(\bm{k},\varepsilon)$.
Especially, one can find from Eq. (\ref{eq:L=1}) that Eq. (\ref{eq:eps=lambda})
has the solution $\varepsilon=\varepsilon_{n}(\bm{k})$
but does not have $\varepsilon=\varepsilon_{m}(\bm{k})$ ($m\neq n$),
which is consistent with Eq. (\ref{eq:ineq-lambda}) since
\begin{equation}
\varepsilon_{+}(\bm{k})=\lambda_{+}(\bm{k},\varepsilon_{+}(\bm{k}))
\geq
\lambda_{-}(\bm{k},\varepsilon_{-}(\bm{k}))=\varepsilon_{-}(\bm{k})
.
\end{equation}
From $L_{n,n}(\bm{k})=1$, we also find that 
the derivative of $\lambda_{n}(\bm{k},\varepsilon)$ with respect to $\varepsilon$
is related to $\cosh\chi_{n}(\bm{k})$ as Eq. (\ref{eq:cosh=lambda}).
The derivation of the relation between
$\bm{v}_{n}(\bm{k})$ and $\bm{w}_{n}(\bm{k},\varepsilon)$ is basically the same as mentioned above.
Then we find the Eq. (\ref{eq:v_w}) and Eq. (\ref{eq:eps=lambda})
has the solution $\varepsilon=-\varepsilon_{n}(-\bm{k})$
but does not have $\varepsilon=-\varepsilon_{m}(-\bm{k})$ ($m\neq n$).

\section{Symmetry}
\label{subsec:symmetry}
In this Appendix, we show systematically 
how the symmetries of a system put some constraint on the form of $\bm{R}(\bm{k},\varepsilon)$. 
In general, a symmetry operation is either described by a unitary or antiunitary
transformation, which are separately discussed. 
As a former example we show that the SU(2) symmetry of a bosonic BdG Hamiltonian of antiferro-magnons
is reduced to the SU(2) symmetry of an effective Hamiltonian $H_{\mathrm{eff}}(\bm{k},\varepsilon)$ and gives $\bm{R}(\bm{k},\varepsilon)=0$. 
As a latter example, it is confirmed that
an effective time-reversal symmetry operation actually flips this pseudo-spin. 

\subsubsection{Unitary transformation}
Suppose that the spin-wave Hamiltonian $\hat {\cal H}_{\rm sw}$ 
in Eq.(\ref{eq:ham_sw}) is invariant under 
the similarity transformation by a unitary operator, $\hat{\mathcal{U}}$, 
\begin{equation}
\hat{\mathcal{U}} \hat {\cal H}_{\rm sw} \hat{\mathcal{U}}^{\dagger}=\hat {\cal H}_{\rm sw}. 
\label{eq:unitarytr}
\end{equation}
We introduce a $2\times2$ unitary matrix, $U$, 
that represents this unitary transformation which should fulfill: 
\begin{equation}
\hat{\mathcal{U}}
\begin{pmatrix}
\hat{a}(\bm{k})\\
\hat{b}(\bm{k})
\end{pmatrix}
\hat{\mathcal{U}}^{\dagger}
=
U^{\dagger}
\begin{pmatrix}
\hat{a}(\bm{k}')\\
\hat{b}(\bm{k}')
\end{pmatrix}
.
\label{eq:sym-unitary}
\end{equation}
Accordingly, the symmetry operation for 
$\hat{\Phi}(\bm{k})=(\hat{a}(\bm{k}),\hat{b}(\bm{k}),\hat{a}^{\dagger}(-\bm{k}),\hat{b}^{\dagger}(-\bm{k}))^{T}$ 
is given as
\begin{equation}
\hat{\mathcal{U}}
\hat{\Phi}(\bm{k})
\hat{\mathcal{U}}^{\dagger}
=
\begin{pmatrix}
U^{\dagger} &\\
& U^{T}
\end{pmatrix}
\hat{\Phi}(\bm{k}')
.
\end{equation}
For such $U$, the elements of $H_{\mathrm{BdG}}(\bm{k})$ satisfy
\begin{align}
&U\Xi(\bm{k})U^{\dagger}
=
\Xi(\bm{k}')
,
\label{eq:sym-Xi}
\\
&U\Delta(\bm{k})U^{T}
=
\Delta(\bm{k}')
.
\label{eq:sym-Delta}
\end{align}
From Eqs. (\ref{eq:heff}), (\ref{eq:sym-Xi}) and (\ref{eq:sym-Delta}),
we find that when the system has the symmetry represented by Eqs.(\ref{eq:unitarytr}), and (\ref{eq:sym-unitary}),
the effective Hamiltonian $H_{\rm eff}(\bm{k},\varepsilon)$ 
also has the symmetry described by the following relation:
\begin{equation}
U
H_{\rm eff}(\bm{k},\varepsilon)
U^{\dagger}
=
H_{\rm eff}(\bm{k}',\varepsilon)
\label{eq:sym-H_eff}
.
\end{equation}
\par
We now consider an SU(2) symmetry of magnon systems the authors introduced in Ref. \cite{kawano2019-1} as an example. 
The corresponding SU(2) algebra 
in the space spanned by 
$\hat{\Phi}(\bm{k})$ is given by the $4\times 4$ matrices,
\begin{equation}
J^{x}=
\tau^{0}\otimes\sigma^{x}
,
\hspace{10pt}
J^{y}=
\tau^{z}\otimes\sigma^{y}
,
\hspace{10pt}
J^{z}=
\tau^{z}\otimes\sigma^{z}
,
\end{equation}
and one can see that these matrices satisfy the SU(2) commutation relation 
$[J^{\mu}, J^{\nu}]=2i\epsilon^{\mu\nu\rho}J^{\rho}$ ($\mu,\nu,\rho=x,y,z$).
If the bosonic BdG Hamiltonian satisfies, $[H_{\mathrm{BdG}}(\bm{k}),J^{\mu}]=0$ ($\mu=x,y,z$), 
the system has a U(1) symmetry about the $\mu$-axis, 
and such a symmetry puts restriction on magnon bands and magnon spin textures in momentum space \cite{kawano2019-1}.
\par
Let us see how this symmetry is reflected in the effective Hamiltonian. 
The $2\times 2$ unitary matrices $U^{\mu}$ which corresponds to $J^{\mu}$ ($\mu=x,y,z$) can be written as 
\begin{equation}
U^{x}=\sigma^{x}
,
\hspace{10pt}
U^{y}=\sigma^{y}
\hspace{10pt}
,
U^{z}=i\sigma^{z}
.
\end{equation}
Thus, when the system has the U(1) symmetry represented by $J^{\mu}$,
the effective Hamiltonian satisfies the following relation:
\begin{equation}
[H_{\rm eff}(\bm{k},\varepsilon),\sigma^{\mu}]=0, 
\label{eq:U1_H_eff}
\end{equation}
which leads to $R^{\mu}(\bm{k},\varepsilon)=0$. 
Therefore, when the system has the SU(2) symmetry,
Eq. (\ref{eq:U1_H_eff}) holds for all $\mu=x,y,z$,
which gives $\bm{R}(\bm{k},\varepsilon)=\bm{0}$.
This reminds us of the single-orbital electronic systems, 
where the U(1) symmetry is described as $[H(\bm{k}),\sigma^{\mu}]=0$, ($\mu=x,y,z$),
and the SU(2) symmetry leads to the doubly degenerate energy bands.

\subsubsection{Antiunitary transformation}
Next, we consider the symmetry represented by the similarity transformation in terms of
an antiunitary operator, $\hat{\mathcal{T}}$, 
\begin{equation}
\hat{\mathcal{T}} \hat {\cal H}_{\rm sw} \hat{\mathcal{T}}^{\dagger}=\hat {\cal H}_{\rm sw}. 
\label{eq:antiunitarytr}
\end{equation}
where $\hat{\mathcal{T}}$ satisfies $\hat{\mathcal{T}}i\hat{\mathcal{T}}^{-1}=-i$. 
The $2\times2$ unitary matrix, $U$, for this transformation satisfies 
\begin{align}
\hat{\mathcal{T}}
\begin{pmatrix}
\hat{a}(\bm{k})\\
\hat{b}(\bm{k})
\end{pmatrix}
\hat{\mathcal{T}}^{-1}
&=
U^{\dagger}
\begin{pmatrix}
\hat{a}(\bm{k}')\\
\hat{b}(\bm{k}')
\end{pmatrix}
,
\\
\hat{\mathcal{T}}
\begin{pmatrix}
\hat{a}^{\dagger}(\bm{k})\\
\hat{b}^{\dagger}(\bm{k})
\end{pmatrix}
\hat{\mathcal{T}}^{-1}
&=
U^{T}
\begin{pmatrix}
\hat{a}^{\dagger}(\bm{k}')\\
\hat{b}^{\dagger}(\bm{k}')
\end{pmatrix}
,
\end{align}
and together with Eq.(\ref{eq:antiunitarytr}), 
the elements of $H_{\mathrm{BdG}}(\bm{k})$ fulfills 
\begin{equation}
UK\Xi(\bm{k})K^{-1}U^{\dagger}
=
\Xi(\bm{k}')
,
\end{equation}
\begin{equation}
UK\Delta(\bm{k})K^{-1}U^{T}
=
\Delta(\bm{k}')
,
\end{equation}
where $K$ is the complex conjugation operator satisfying $KiK^{-1}=-i$.
Accordingly, from Eqs. (\ref{eq:heff}), (\ref{eq:sym-Xi}) and (\ref{eq:sym-Delta}), 
the effective Hamiltonian satisfies
\begin{equation}
UKH_{\rm eff}(\bm{k},\varepsilon)K^{-1}U^{\dagger}
=
H_{\rm eff}(\bm{k}',\varepsilon)
.
\end{equation}
\par
As an example, we consider the effective time-reversal symmetry for antiferro-magnons 
introduced in Refs. \cite{kondo2019-1,kondo2019-2}, 
which leads to the Kramers pair of magnons and the $\mathbb{Z}_{2}$ topological invariant 
for this symmetry corresponds to a class AII \cite{schnyder2008}.
This fact implies that such an effective time-reversal symmetry flips the pseudo-spin of antiferro-magnons,
as the time-reversal symmetry for electrons does to the electronic spins. 
The effective time-reversal symmetry operation is represented by the $4\times 4$ matrices as 
\begin{equation}
\tilde{\Theta}_{tr}H_{\mathrm{BdG}}(-\bm{k})\tilde{\Theta}_{tr}^{-1}=H_{\mathrm{BdG}}(\bm{k})
,
\end{equation}
with $\tilde{\Theta}_{tr}=iJ^{y}K$. 
The corresponding symmetry for a $2\times 2$ effective Hamiltonian is given as
\begin{equation}
\Theta_{tr}H_{\rm eff}(-\bm{k},\varepsilon)\Theta_{tr}^{-1}=H_{\rm eff}(\bm{k},\varepsilon)
,
\end{equation}
with $\Theta_{tr}=i\sigma^{y}K$, which leads to the following relation:
\begin{equation}
\bm{R}(-\bm{k},\varepsilon)=-\bm{R}(\bm{k},\varepsilon)
.
\label{eq:R_flip}
\end{equation}
In the electronic systems, the true time-reversal operator takes the same form as the above $\Theta_{tr}$,
which then gives $\bm{R}(-\bm{k})=-\bm{R}(\bm{k})$ for $\bm{R}(\bm{k})$ in Eqs.(\ref{eq:H_electron}) and (\ref{eq:S_el}).
We thus confirm that $\bm \sigma$ in the effective Hamiltonian of magnons 
actually serves as a pseudo-spin degrees of freedom of antiferro-magnons.
\section{Other spin textures in the KH$\Gamma$ model}
\label{sec:app_khg}
When $\Gamma=0$, the classical energy in Eq. (\ref{eq:cgs}) is reduced to the form 
\begin{equation}
E_{\mathrm{cgs}}=
-(3J+2K)N_{c}S^{2}\cos2\zeta-2(\bm{h}\cdot\bm{e}_{u})N_{c}S\sin\zeta
,
\end{equation}
and one can immediately see that $\bm{e}_{u}=\bm{h}/h$.
The canting angle is given as $\zeta=\arcsin\big(h/(2(3J+2K)S)\big)$.
The lowest energy of $E_{\mathrm{cgs}}$ has a 
degeneracy in terms of rotating $\bm{e}_{s}$ about the $\bm{e}_{u}$-axis, which is described by the angle $\eta=0\sim 2\pi$
(see Fig.\ref{fig3}(b)). 
When $K\neq0$, this degeneracy is accidental since the system has no spin-rotational symmetry about $\bm e_u$. 
This degeneracy is lifted quantum mechanically via the order-by-disorder mechanism \cite{villain1980,shender1982,henley1989}. 
The angle $\eta$ is thus determined by minimizing the following zero-point fluctuation energy density, 
\begin{equation}
\varepsilon_{\mathrm{zero}}(\eta)=
\frac{1}{2}\int_{\mathrm{BZ}}\frac{d^{2}\bm{k}}{(2\pi)^{2}}\
\sum_{n=\pm}\varepsilon_{n}(\bm{k})
,
\end{equation}
where we get the energy bands $\varepsilon_{\pm}(\bm{k})$ numerically,
and the integral runs over the Brillouin zone.
Even if $\Gamma\neq0$, the classical ground state becomes degenerates
when the magnetic field is parallel to the [111]-direction, $\bm{h}/h=(1,1,1)/\sqrt{3}$,
and we need to consider the zero-point fluctuation to determine the direction of $\bm{e}_{s}$.
\par
\begin{figure*}[tbp]
\includegraphics[width=180mm]{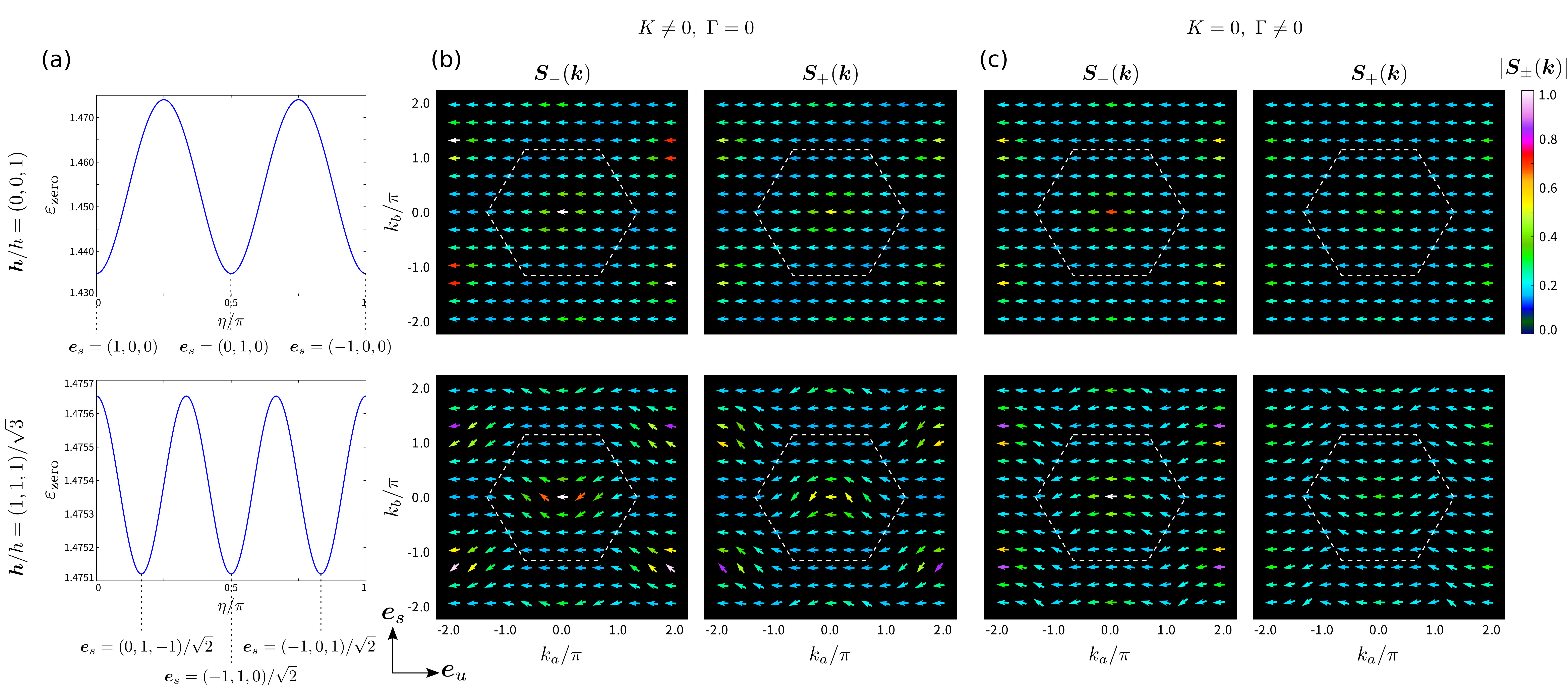}
\caption{(a) Zero-point energy density as a function of $\eta$.
$\bm{e}_{s}$ is parametrized as $\bm{e}_{s}=(\cos\eta,\sin\eta,0)$ for $\bm{h}/h=(0,0,1)$
and $\bm{e}_{s}=(\sqrt{1/6}\cos\eta-\sqrt{1/2}\sin\eta,\sqrt{1/6}\cos\eta+\sqrt{1/2}\sin\eta,-\sqrt{2/3}\cos\eta)$
for $\bm{h}/h=(1,1,1)\sqrt{3}$.
(b,c) Spin carried by magnons, $\bm{S}_{\pm}(\bm{k})$,
for (b) $\Theta=\pi/3$ and $\Gamma=0$ and (c) $\Theta=0$ and $\Gamma=-0.1$
with a magnetic field $h=0.5$, $\bm{h}/h=(0,0,1)$ and $(1,1,1)/\sqrt{3}$.
}
\label{figapp}
\end{figure*}

Figure \ref{figapp}(a) shows $\varepsilon_{\mathrm{zero}}(\eta)$
as a function of $\eta$ for two different choices of field directions.
For the magnetic field in the $z$-direction, $\bm{h}/h=(0,0,1)$,
$\bm{e}_{s}=\pm\bm{e}_{x}$ and $\pm\bm{e}_{y}$ are selected by the zero-point fluctuation.
The similar behavior of $\varepsilon_{\rm zero}(\eta)$ is also observed
when $\bm{h}/h$ lies in the $xy$-, $yz$-, and $zx$-plane,
which leads to $\bm{e}_{s}=\pm\bm{e}_{z}$, $\bm{e}_{s}=\pm\bm{e}_{x}$ and $\bm{e}_{s}=\pm\bm{e}_{y}$, respectively.
For the magnetic field in the [111] direction, $\bm{h}/h=(1,1,1)/\sqrt{3}$, which is perpendicular to the honeycomb plane,
$\bm{e}_{s}=(0,1,-1)/\sqrt{2}$ is selected, for example.
\par
The direction of the uniform magnetization $\bm e_u$ is parallel to $\bm h$, 
and the classical degeneracy is such that the $\bm e_s$ rotates around $\bm h$. 
When $\bm{h}$ lies in the $xy$-, $yz$-, and $zx$-plane, the zero-point fluctuation selects
$\bm{e}_{s}$ to either of the three links, $\mu=x,y,z$.
This configuration maximally gains the zero point fluctuation energy since the fluctaution of magnons 
along the other two links could equivalently contribute to the zero point flutuation energy.
In fact, when $\bm{h}/h=(1,0,0)$, $\bm{e}_s$ points in the $y$- or $z$-directions which are perpendicular to the 
$y$- and $z$-links, respectively.
However, for $\bm{h}/h=(1,1,1)/\sqrt{3}$ (perpendicular to the honeycomb plane),
one cannot take $\bm e_s$ in that manner.
\par
When $\bm{e}_s$ is perpendicular to either of the $x,y,z$-links, we find $\mathrm{Im}\bm{Q}=0$. 
For example, when $\bm e_s \parallel \bm x$, 
\begin{equation}
\mathrm{Im}\bm{Q}=
KS\sin\zeta\sin2\eta 
\begin{pmatrix}
\bm e_s\cdot(\bm e_s\times \bm e_u) \\
0\\
0
\end{pmatrix}
=\bm 0
,
\label{eq:ImQ-hx}
\end{equation}
and this immediately gives $R^{z}(\bm{k},\varepsilon)=0$ from Eqs.(\ref{eq:Rz-kitaev}) and (\ref{eq:condition_Q}).
Therefore, when $\Gamma=0$ and $\bm{h}$ lies in the $xy$-, $yz$- and $zx$-plane, a type-(ii) spin texture realizes.
One can see $\mathrm{Im}\bm{Q}=0$ line in Fig. \ref{fig4} (c).
\par
We show in Figs.\ref{figapp}(b) and (c) several cases of spin textures with and without $K$ or $\Gamma$.
As discussed above, for $\Gamma=0$, the magnetic moments point in the direction perpendicular to the $y$ or $z$-axis for 
$\bm{h}/h=(0,0,1)$, in which case $R^{z}(\bm{k},\varepsilon)=0$ and type-(ii) is realized.
Otherwise, we basically find type-(iii) while the variation of the direction of $\bm{S}_\pm(\bm k)$ is not necessarily rich 
for the cases presented here.
It can be explained by the field-direction dependence of $|\mathrm{Im}\bm{Q}|$ as shown in Fig. \ref{fig4}:
When $K=0$ or $\Gamma=0$, $|\mathrm{Im}\bm{Q}|$ is smaller than that of $K\neq0$ and $\Gamma\neq0$.
Therefore, one needs to carefully examine whether the value of $\bm Q$ in Eq.(\ref{eq:condition_Q})
determined solely by the field direction (minimizing the ground state energy) has a large enough value to
contribute to the variation of $R^z(\varepsilon,\bm k)$.
%
%
\section{Analytical Expression of $\bf \it P$ and $\bf \it Q$ of the KH$\Gamma$ model}
\label{sec:app_pq}
We show the details of the bosonic BdG Hamiltionian of the KH$\Gamma$ model.
We choose the local right-handed coordinate basis, $\bm{X}_{A/B}$ and $\bm{Y}_{A/B}$ as
\begin{align}
& \bm{X}_{A}
=
\bm{e}_{u}\cos\zeta-\bm{e}_{s}\sin\zeta
,
\hspace{20pt}
\bm{X}_{B}
=
-\bm{e}_{u}\cos\zeta-\bm{e}_{s}\sin\zeta
,
\nonumber\\
& \bm{Y}_{\mathrm{A}}=\bm{Y}_{\mathrm{B}}=
\bm{e}_{s}\times\bm{e}_{u}
.
\end{align}
The Holstein-Primakoff transformation for a bilinear spin interaction
$\hat{S}_{i\in A}^{\mu}\hat{S}_{j\in B}^{\nu}$
($\mu,\nu=x,y,z$) is given by
\begin{align}
\hat{S}_{i\in A}^{\mu}\hat{S}_{j\in B}^{\nu}
&=
\left\{
S^{2}-S(\hat{a}_{i}^{\dagger}\hat{a}_{i}+\hat{b}_{j}^{\dagger}\hat{b}_{j})
\right\}
M_{A}^{\mu}M_{B}^{\nu}
\nonumber\\
&\hspace{20pt}
+\left(
C_{AB}^{\mu\nu}\hat{a}_{i}^{\dagger}\hat{b}_{j}^{\dagger}
+
I_{AB}^{\mu\nu}\hat{a}_{i}^{\dagger}\hat{a}_{j}
+
\mathrm{h.c.}
\right)
,
\end{align}
where
\begin{align}
C_{AB}^{\mu\nu}
&=
\frac{S}{2}
\{
X_{A}^{\mu}X_{B}^{\nu}-Y_{A}^{\mu}Y_{B}^{\nu}
+i
(X_{A}^{\mu}Y_{B}^{\nu}+Y_{A}^{\mu}X_{B}^{\nu})
\}
,
\label{eq:C_munu}
\\
I_{AB}^{\mu\nu}
&=
\frac{S}{2}
\{
X_{A}^{\mu}X_{B}^{\nu}+Y_{A}^{\mu}Y_{B}^{\nu}
-i
(X_{A}^{\mu}Y_{B}^{\nu}-Y_{A}^{\mu}X_{B}^{\nu})
\}
.
\label{eq:I_munu}
\end{align}
$\bm{P}=(P_{x},P_{y},P_{z})^{T}$ and $\bm{Q}=(Q_{x},Q_{y},Q_{z})^{T}$ in
Eqs. (\ref{eq:Xi_x_KHGamma}), (\ref{eq:Xi_y_KHGamma}), (\ref{eq:Delta_x_KHGamma}) and (\ref{eq:Delta_y_KHGamma})
are expressed as
\begin{align}
\bm{P}
&=
JS\sin^{2}\zeta
\begin{pmatrix}
1\\
1\\
1
\end{pmatrix}
+
2K
\begin{pmatrix}
I_{AB}^{xx}\\
I_{AB}^{yy}\\
I_{AB}^{zz}
\end{pmatrix}
+
\Gamma
\begin{pmatrix}
I_{AB}^{yz}+I_{AB}^{zy}\\
I_{AB}^{zx}+I_{AB}^{xz}\\
I_{AB}^{xy}+I_{AB}^{yx}
\end{pmatrix}
,
\\
\bm{Q}
&=
-JS\cos^{2}\zeta
\begin{pmatrix}
1\\
1\\
1
\end{pmatrix}
+
2K
\begin{pmatrix}
C_{AB}^{xx}\\
C_{AB}^{yy}\\
C_{AB}^{zz}
\end{pmatrix}
+
\Gamma
\begin{pmatrix}
C_{AB}^{yz}+C_{AB}^{zy}\\
C_{AB}^{zx}+C_{AB}^{xz}\\
C_{AB}^{xy}+C_{AB}^{yx}
\end{pmatrix}
.
\end{align}
From Eq.(\ref{eq:condition_Q}), we need to know the explicit form of $\mathrm{Re}\bm{Q}$ and $\mathrm{Im}\bm{Q}$.
The latter is presente in Eq.(\ref{eq:imq}) and the former is given as
\begin{align}
\mathrm{Re}\bm{Q}&=
-JS\cos^{2}\zeta
\begin{pmatrix}
1\\
1\\
1
\end{pmatrix}
-
KS\
\mathrm{Re}\bm{Q}_{K}
-
\Gamma S\
\mathrm{Re}\bm{Q}_{\Gamma}
,
\end{align}
where
\begin{align}
\mathrm{Re}\bm{Q}_{K}&=
\begin{pmatrix}
(e_{u}^{x})^{2}\cos^{2}\zeta-(e_{s}^{x})\sin^{2}\zeta+\{(\bm{e}_{s}\times\bm{e}_{u})^{x}\}^{2}\\
(e_{u}^{y})^{2}\cos^{2}\zeta-(e_{s}^{y})\sin^{2}\zeta+\{(\bm{e}_{s}\times\bm{e}_{u})^{y}\}^{2}\\
(e_{u}^{z})^{2}\cos^{2}\zeta-(e_{s}^{z})\sin^{2}\zeta+\{(\bm{e}_{s}\times\bm{e}_{u})^{z}\}^{2}
\end{pmatrix}
\nonumber \\
\mathrm{Re}\bm{Q}_{\Gamma}&=
\begin{pmatrix}
e_{u}^{y}e_{u}^{z}\cos^{2}\zeta-e_{s}^{y}e_{s}^{z}\sin^{2}\zeta+(\bm{e}_{s}\times\bm{e}_{u})^{y}(\bm{e}_{s}\times\bm{e}_{u})^{z}\\
e_{u}^{z}e_{u}^{x}\cos^{2}\zeta-e_{s}^{z}e_{s}^{x}\sin^{2}\zeta+(\bm{e}_{s}\times\bm{e}_{u})^{z}(\bm{e}_{s}\times\bm{e}_{u})^{x}\\
e_{u}^{x}e_{u}^{y}\cos^{2}\zeta-e_{s}^{x}e_{s}^{y}\sin^{2}\zeta+(\bm{e}_{s}\times\bm{e}_{u})^{x}(\bm{e}_{s}\times\bm{e}_{u})^{y}
\end{pmatrix}
\end{align}
which is apparently nonzero.
Therefore, examining $\mathrm{Im}\bm{Q}$ is important to clarify the behavior of $R^{z}(\bm{k},\varepsilon)$ 
as performed in the main text. 
\section{Details of the BdG Hamiltonian of the honeycomb antiferromagnet}
\label{app:honeycomb}
We show the parameter values of the bosonic BdG Hamiltonian of the honeycomb antiferromagnet in 
\S.\ref{sec:honeycomb}. 
When we consider the z-AFM phase, the bosonic BdG Hamiltonian given in 
Eq.(\ref{eq:bdg_honey}) consists of parameters 
\begin{align}
&\Xi_{0}(\bm{k})=
3(J_{1}+G_{1})S-6(J_{2}+G_{2})S
\nonumber \\
&\hspace{20pt}
+
2\mathrm{Re}
\left[
(J_{2}-G_{2}+iD)S
\sum_{\bm{\delta}'}
\mathrm{e}^{i\bm{k}\cdot\bm{\delta}'}
\right]
,
\\
&\Delta_{x}(\bm{k})=
-(J_{1}-G_{1})S
\sum_{\mu}
\cos\bm{k}\cdot\bm{\delta}_{\mu}
,
\\
&\Delta_{y}(\bm{k})=
(J_{1}-G_{1})S
\sum_{\mu}
\sin\bm{k}\cdot\bm{\delta}_{\mu}
.
\end{align}
\par
For the xy-AFM phase with $\bm h\parallel \bm a$, 
the parameters in Eq.(\ref{eq:bdg_honey}) are given as 
\begin{align}
\Xi_{0}(\bm{k})&=
3(J_{1}-G_{1})S\cos2\zeta-6(J_{2}-G_{2})S
\nonumber \\
&\hspace{20pt}
+h\sin\zeta
+2J_{2}S
\mathrm{Re}
\left[
\sum_{\bm{\delta}'}
\mathrm{e}^{i\bm{k}\cdot\bm{\delta}'}
\right]
,
\label{eq:honeycomb-xi0}
\\
\Xi_{x}(\bm{k})&=
(J_{1}\sin^{2}\zeta+G_{1}\cos^{2}\zeta)S
\sum_{\mu}
\cos\bm{k}\cdot\bm{\delta}_{\mu}
,
\\
\Xi_{y}(\bm{k})&=
-(J_{1}\sin^{2}\zeta+G_{1}\cos^{2}\zeta)S
\sum_{\mu}
\sin\bm{k}\cdot\bm{\delta}_{\mu}
,
\\
\Delta_{0}(\bm{k})&=
-2G_{2}S
\mathrm{Re}
\left[
\sum_{\bm{\delta}'}
\mathrm{e}^{i\bm{k}\cdot\bm{\delta}'}
\right]
,
\\
\Delta_{x}(\bm{k})&=
-(J_{1}\cos^{2}\zeta+G_{1}\sin^{2}\zeta)S
\sum_{\mu}
\cos\bm{k}\cdot\bm{\delta}_{\mu}
,
\\
\Delta_{y}(\bm{k})&=
(J_{1}\cos^{2}\zeta+G_{1}\sin^{2}\zeta)S
\sum_{\mu}
\sin\bm{k}\cdot\bm{\delta}_{\mu}
,
\label{eq:honeycomb-deltay}
\end{align}
which do not include $D$.
\par
When $\bm h \parallel \bm c$ is applied to the xy-AFM, the parameters become 
\begin{align}
\Xi_{0}(\bm{k})&=
3(J_{1}+2G_{2})S\cos2\zeta-3(G_{1}+2J_{2})S
\nonumber \\
&\hspace{20pt}
+h\sin\zeta
+2(J_{2}-G_{2}\sin^{2}\zeta)S
\mathrm{Re}
\left[
\sum_{\bm{\delta}'}
\mathrm{e}^{i\bm{k}\cdot\bm{\delta}'}
\right]
,
\label{eq:honeycomb-xi0-2}
\\
\Xi_{x}(\bm{k})&=
(J_{1}\sin^{2}\zeta-G_{1})S
\sum_{\mu}
\cos\bm{k}\cdot\bm{\delta}_{\mu}
,
\\
\Xi_{y}(\bm{k})&=
-(J_{1}\sin^{2}\zeta-G_{1})S
\sum_{\mu}
\sin\bm{k}\cdot\bm{\delta}_{\mu}
,
\\
\Xi_{z}(\bm{k})&=
-2DS\sin\zeta
\mathrm{Im}
\left[
\sum_{\bm{\delta}'}
\mathrm{e}^{i\bm{k}\cdot\bm{\delta}'}
\right]
,
\label{eq:honeycomb-xiz}
\\
\Delta_{0}(\bm{k})&=
-2G_{2}S\cos^{2}\zeta
\mathrm{Re}
\left[
\sum_{\bm{\delta}'}
\mathrm{e}^{i\bm{k}\cdot\bm{\delta}'}
\right]
,
\\
\Delta_{x}(\bm{k})&=
-J_{1}\cos^{2}\zeta
\sum_{\mu}
\cos\bm{k}\cdot\bm{\delta}_{\mu}
\\
\Delta_{y}(\bm{k})&=
J_{1}\cos^{2}\zeta
\sum_{\mu}
\sin\bm{k}\cdot\bm{\delta}_{\mu}
.
\label{eq:honeycomb-deltay-2}
\end{align}
which gives type-(iii) spin texture. 


\bibliography{biblio}
\bibliographystyle{apsrev4-1_mk}

\end{document}